\documentclass[twocolumn,prl,superscriptaddress,longbibliography]{revtex4-1} 
\usepackage[english]{babel}
\usepackage{graphicx}
\usepackage{amssymb}
\usepackage{amsmath}
\usepackage{color}
\usepackage{comment}
\usepackage{bm}
\usepackage[space]{grffile}   

\usepackage{hyperref}

\newcommand*{\beq}{\begin{equation}}
\newcommand*{\eeq}{\end{equation}}

\usepackage{mathtools}

\DeclarePairedDelimiter\ket{\lvert}{\rangle}

\newcommand{\E}{\mathcal{E}}
\newcommand{\up}{\uparrow}
\newcommand{\down}{\downarrow}

\hypersetup{%
   pdfpagemode=None, 
   pdfstartpage=1,
   pdfmenubar=true,
   pdftoolbar=true,
   colorlinks = true,
   linkcolor=blue,
   citecolor=blue,
   urlcolor=blue,
   bookmarksopen=false
 }

\begin{document}
\title{Crossover between few and many fermions in a harmonic trap}							%

\author{Tomasz Grining}
\affiliation{Faculty of Chemistry, University of Warsaw, Pasteura 1, 02-093 Warsaw, Poland}
\author{Micha\l~Tomza}
\affiliation{ICFO-Institut de Ciencies Fotoniques, The Barcelona Institute of Science and Technology, Av.~Carl Friedrich Gauss 3, 08860 Castelldefels (Barcelona), Spain}
\author{Micha\l~Lesiuk}
\affiliation{Faculty of Chemistry, University of Warsaw, Pasteura 1, 02-093 Warsaw, Poland}   
\author{Micha\l~Przybytek}
\affiliation{Faculty of Chemistry, University of Warsaw, Pasteura 1, 02-093 Warsaw, Poland}
\author{Monika~Musia{\l}}
\affiliation{Institute of Chemistry, University of Silesia, Szkolna 9, 40-006 Katowice, Poland}
\author{Robert Moszynski}
\affiliation{Faculty of Chemistry, University of Warsaw, Pasteura 1, 02-093 Warsaw, Poland}
\author{Maciej Lewenstein}
\affiliation{ICFO-Institut de Ciencies Fotoniques, The Barcelona Institute of Science and Technology, Av.~Carl Friedrich Gauss 3, 08860 Castelldefels (Barcelona), Spain}
\affiliation{ICREA-Instituci\'o Catalana de Recerca i Estudis Avan\c{c}ats, 08010 Barcelona, Spain}
\author{Pietro Massignan}
\affiliation{ICFO-Institut de Ciencies Fotoniques, The Barcelona Institute of Science and Technology, Av.~Carl Friedrich Gauss 3, 08860 Castelldefels (Barcelona), Spain}

\begin{abstract}
The properties of a balanced two-component Fermi gas in a one-dimensional harmonic trap are studied by means of the coupled cluster method. For few fermions we recover the results of exact diagonalization, yet with this method we are able to study much larger systems. We compute the energy, the chemical potential, the pairing gap, and the density profile of the trapped clouds, smoothly mapping the crossover between the few-body and many-body limits. The energy is found to converge surprisingly rapidly to the many-body result for every value of the interaction strength. Many more particles are instead needed to give rise to the non-analytic behavior of the pairing gap, and to smoothen the pronounced even-odd oscillations of the chemical potential induced by the shell structure of the trap.
\end{abstract}

\date{\today}


\maketitle

Ultracold gases are ideal systems for engineering highly non-trivial states of matter. They allow one to prepare, manipulate, and measure with great accuracy strongly correlated quantum systems  \cite{Lewenstein2012,Bloch2008,HouchesVol94}. Thus, they provide a perfect playground for classical simulations, and can accurately serve as quantum simulators \cite{Lewenstein2007}.
One-dimensional (1D) systems are particularly fascinating because of the important role played by quantum fluctuations \cite{Giamarchi}. 
Experimental  studies of strongly correlated ultracold atomic gases started with the seminal works on the Tonks-Girardeau gas \cite{Tilman1D,Weiss1D,Belen}. More recently the super Tonks-Girardeau regime has been achieved \cite{Haller2009}, and the first experiments with fermions with a tunable spin  have been launched \cite{Pagano2014}.
For this Rapid Communication, particularly relevant are experiments on two-component Fermi gas performed in 1D harmonic traps containing very few atoms, where the number of spin-up and spin-down atoms may be fully, and separately, controlled \cite{Serwane2011,Zurn2012,Wenz2013,Zurn2013,Murmann2015}.
The theoretical studies of 1D Fermi gases in the many-body regime have a long tradition  \cite{Recati2003,Juillet2004,Astrakharchik2004,Tokatly2004,Fuchs2004,Carusotto2004,Astrakharchik2005,Orso2007,Hu2007,ColomeTatche2008,Casula2008,Soeffing2011}. More recently, several papers have addressed in detail the few-body case, and its evolution towards the many-body regime \cite{Gharashi2013,Sowinski2013,Astrakharchik2013,Deuretzbacher2014,Volosniev2014,Lindgren2014,Levinsen2015,DAmico2015,Berger2015,Sowinski2015}.

In order to address increasingly more elaborate experimental findings, the efficient numerical treatment of many-body quantum systems stands as one of the great challenges of modern physics. Despite enormous progress and the development of many powerful approaches (cf., e.g., density functional theory \cite{Dreizler2012}, exact diagonalization \cite{Sandvik2010,alps}, quantum Monte Carlo~(QMC)~\cite{Casula2008,alps,VanHoucke2010}, density matrix renormalization group~(DMRG) and tensor network states 
\cite{schollwock1,schollwock2})
, new numerical approaches that are able to investigate hitherto unexplored phenomena are always more than welcome.

In this Rapid Communication, we study a balanced two-component Fermi gas in a one-dimensional harmonic trap by means of a quantum chemistry approach---the coupled cluster (CC)  method \cite{Coester1958,Cizek1966,*Cizek1969,Cizek1971,Bartlett1981,*Bartlett1989,Bishop1991,Paldus1999,Musial2007,Lyakh2012}. 
In condensed matter, this method has up to now been successfully applied to spin-1/2 lattice models in 1D and 2D (see Refs. \cite{Bishop1994,Bishop2011} and references therein), and to trapped ultracold bosons~\cite{Cederbaum2006,Alon2006}. However, the CC method is ideally suited to study fermionic systems, where the number of occupied orbitals grows at least as fast as the number of particles in the system, even in the absence of interactions.
CC recovers results known from exact diagonalization \cite{DAmico2015} and a path integral approach \cite{Hofmann2015}, but it also allows one to study much larger systems (up to $\simeq 80$ particles).
QMC methods permit one to look at even bigger clouds \cite{Casula2008}; ground state properties may be studied by means of the very accurate diffusion QMC, while finite temperatures may be considered via path-integral MC.
Finally, CC compares well with the state-of-art 
DMRG calculations~\cite{Soeffing2011}, with the advantage however of being a method explicitly working in continuous space rather than in a lattice. 

Here we compute with high precision the energy, the chemical potential, the pairing gap and the density profile of the fermionic gas as a function of the number of particles and of the interaction strength. While the ground state energy of the system converges astonishingly rapidly to the many-body result for any interaction strength, 
the other quantities depend more sensitively on system size and interaction strength. In particular, strong even/odd oscillations in the chemical potential of the trapped gas persist up to a very large number of particles, and the non-analytic behavior of the BCS pairing gap emerges only relatively slowly with the system size.


\begin{figure}
\centering
\includegraphics[width=\columnwidth]{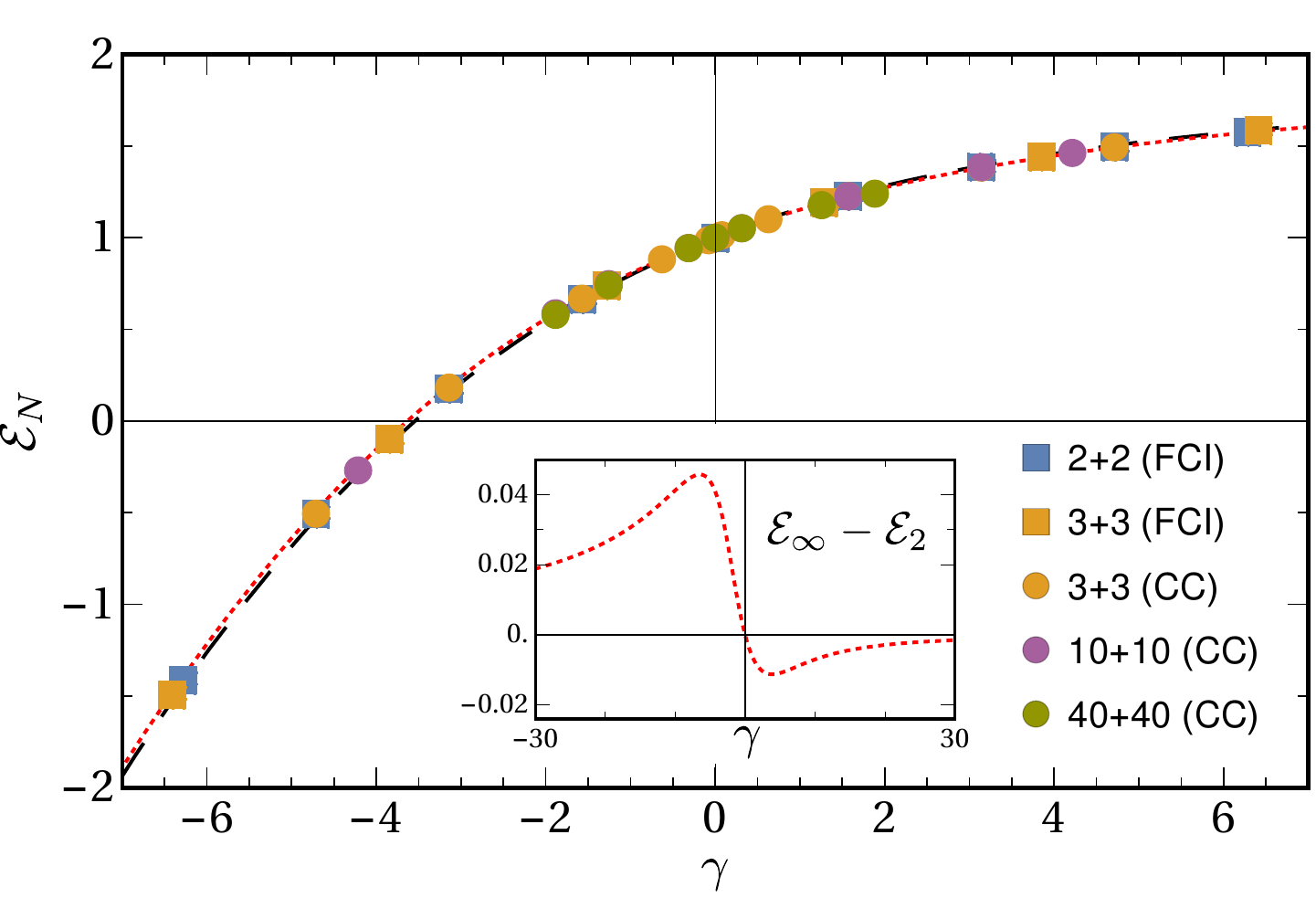}
\caption{Rescaled energies $\E_N=E_N/E_N^{(0)}$ for various number of fermions, as a function of the interaction strength $\gamma$. The squares and circles are respectively the results of the FCI and CC calculations. The black dashed line is the analytic two-body result $\E_2$ for 1+1 particles \cite{Busch1998}, and the red dotted line is the thermodynamic result $\E_\infty$ obtained from the GY+LDA approach~\cite{Astrakharchik2004}. The inset shows the difference $\E_\infty-\E_2$ vs.\ $\gamma$, which remains surprisingly small for every interaction strength. }
\label{fig:rescaledEnergies}
\end{figure}

{\bf Model.}
We consider a two-component Fermi gas containing $N=N_\up+N_\down$ atoms in a balanced configuration, i.e., with $N_\up=N_\down$. All atoms have the same mass $m$, and they are bound to move along one dimension due to the presence of a strong transverse confinement. Along the axial direction the atoms are further confined by a harmonic potential of frequency $\omega$, and they interact by means of a short-range (contact) interaction of strength $g$. The corresponding Hamiltonian is then
\beq
\mathcal{H}=\sum_{i=1}^N \left(\frac{p_i^2}{2m}+\frac{m\omega^2z_i^2}{2}\right)+g\sum_{i<j}\delta(z_i-z_j).
\eeq
In the absence of interactions, the total energy of the trapped gas is $E_{N}^{(0)}=N^2\hbar\omega/4$, while the Fermi energy (defined as the energy of the first unoccupied level) is $E_F=(N+1)\hbar\omega/2$.
The interaction strength may be suitably parametrized in terms of the dimensionless constant
$\gamma=({\pi g}/{\sqrt{N}})/(\hbar\omega a_z)$, where $a_z = \sqrt{\hbar /m\omega}$ is the harmonic oscillator length. As the interactions are varied, the system continuously evolves from a Tonks-Girardeau (TG) gas of $N$ strongly repulsive fermions (at $\gamma\gg1$) to a Lieb-Liniger (LL) gas of $N/2$ hard-core bosonic dimers (at $\gamma\ll-1$) \cite{Astrakharchik2004}.

{\bf Full configuration interaction (FCI) and CC methods.}
The exact solution of the many-body Schr\"odinger
equation of a fermionic system can always be written as a linear combination of all Slater determinants that may be obtained from a complete set of one-particle basis functions. In a numerical solution, however, only a finite number $n_b$ of functions may generally be taken into account.
The simplest approximation to the exact wave function is the Slater determinant $\ket{\Phi}$
 obtained by solving the mean-field Hartree-Fock (HF) equations. 
The best possible solution in a given finite basis set can be obtained by expanding the wave function in all Slater determinants that may be obtained by replacing (i.e., exciting) one-particle functions in $\ket{\Phi}$, and optimizing variationally the coefficients in the resulting linear combination. This method is termed full configuration interaction (FCI) \cite{Sherrill1999}, or exact diagonalization, and it provides a strict upper bound to the exact ground-state energy.
The computational cost of an FCI calculation scales as $N^2  n_b^{N+2}$ for $n_b\gg N$ \cite{Olsen1988}, which forces a tradeoff between the number of particles in the system and the number of basis functions $n_b$ needed to accurately describe it. At present, we are limited to $n_b \approx 50$ for $N=6$. 
One possible way to overcome this obstacle is to limit the
number of excitations included in the FCI wave function. This leads to considerable savings of computer time, but unfortunately any truncated CI calculation is size-inconsistent,
in the sense that the total energy of two systems, not interacting with each other, is not guaranteed to be the sum of the energies of the two systems.

To overcome the size-consistency problem, the coupled cluster (CC) method was introduced,
first in nuclear physics \cite{Coester1960} and shortly after in quantum chemistry \cite{Cizek1966}.
The CC wave function is given by the exponential Ansatz
\beq
\ket{\Psi} = e^{\widehat T}\ket{\Phi},
\label{ccexpan}
\eeq
where $\widehat T = \widehat T_1 + \widehat T_2 + \ldots + \widehat T_N $ is the sum of all possible excitation operators $\widehat T_k$ replacing $k$ HF one-particle orbitals in the reference Slater determinant with $k$ orbitals which are not present in $\ket \Phi$.
Due to the exponential form of the Ansatz (\ref{ccexpan}), the CC method truncated to single and double excitations effectively includes various triply, quadruply, and higher excited determinants, since it contains, e.g., the products $\widehat{T}_1\widehat{T}_2$ and $\widehat{T}_2^2$.
The CC method including all excitations is obviously equivalent to the FCI method, and its computational cost is equally high. However, truncated CC calculations are by construction size-consistent, and are much less time consuming than FCI with the same basis set size $n_b$.
In the present work we use the coupled cluster method restricted to single, double, and non-iterative triple excitations, CCSD(T), whose computational complexity scales as $n_b^7$ \cite{Raghavachari1989}. The CCSD(T) method is very accurate for many properties of atoms and molecules, and it is now considered the golden standard of quantum chemistry~\cite{Musial2007}. 

We construct our many-body wave function by expanding it on the restricted basis containing the first $n_b$ single-particle eigenfunctions of the 1D harmonic oscillator,
\mbox{$\phi_n(z)=
 H_n(z/a_z)\exp{(-z^2/2a_z^2)}/\sqrt{2^n n! a_z\sqrt{\pi}}$}, with $H_n(.)$ an Hermite polynomial.
A detailed study of the convergence with the type of excitations included in the coupled cluster model and the size of the basis set have been reported elsewhere~\cite{Grining2015}. For the purposes of this work, we simply note that the CCSD(T) method appears to be the best choice for 1D fermionic systems, in terms of the tradeoff between accuracy and computational cost.
As the energy is found to converge to the exact value with a rate $\sim 1/\sqrt{n_b}$, all results presented here are obtained by extrapolating to the limit of infinite basis set with a quadratic interpolation vs.\ $1/\sqrt{n_b}$ passing through the energies obtained for three values of $n_b$ up to 200.
The FCI and CC calculations were performed with customized versions of the HECTOR~\cite{HECTOR} and ACESII codes~\cite{ACESII}, respectively.

\begin{figure}
\centering
\includegraphics[width=\columnwidth]{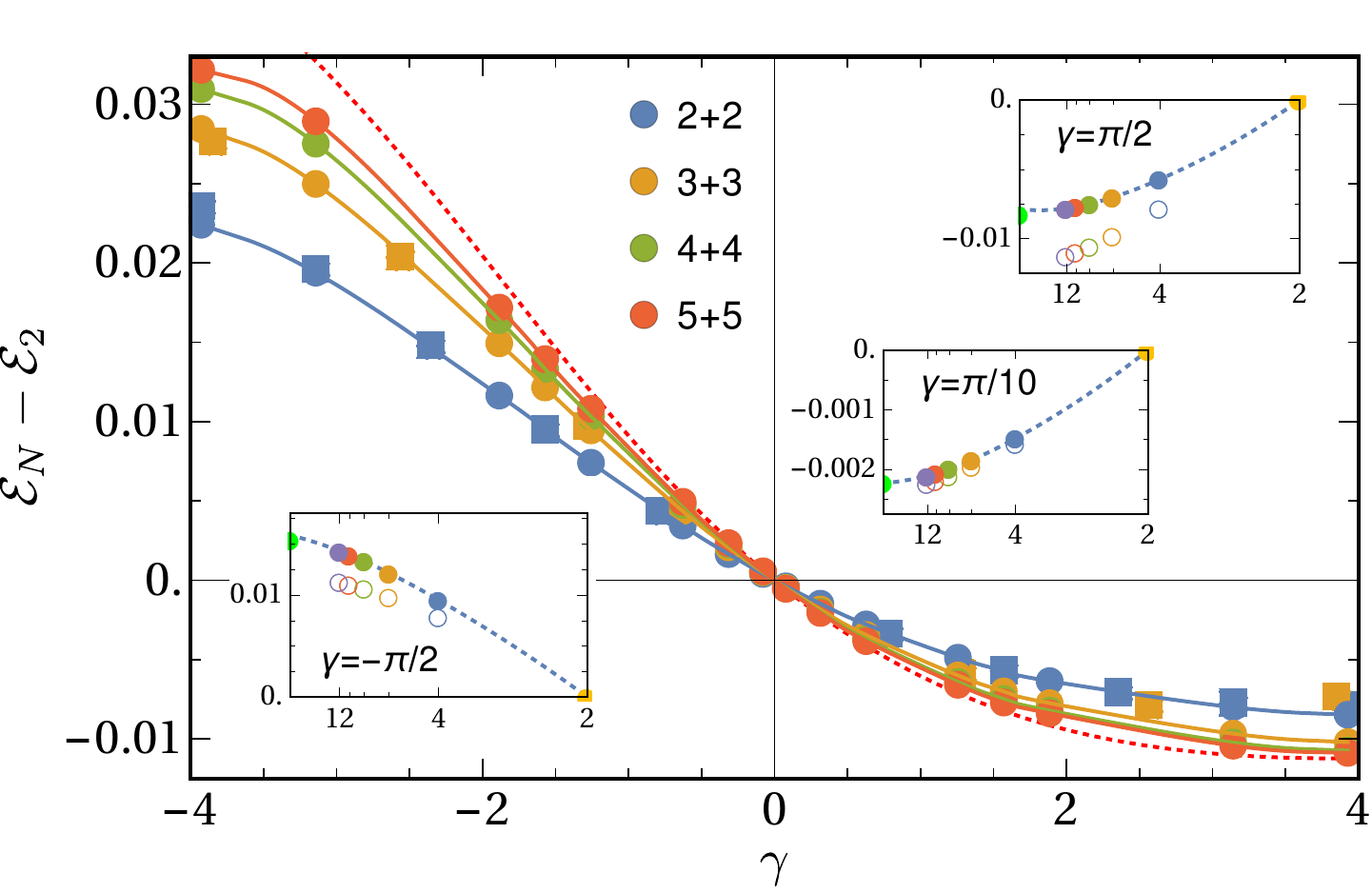}
\caption{Difference between the rescaled few-body energies and the analytical 2-body result; filled circles (squares) are the CC (FCI) results, while the red dotted line is the GY+LDA result.
The insets are vertical cuts across the main figure (i.e., at fixed $\gamma$), plotted as a function of $1/N$. The bright green dots are the expected many-body limit, and the empty circles are the first order perturbation6 approximation,~Eq.~\eqref{enDiffMeanField}. 
}
\label{fig:energyDifferences}
\end{figure}

{\bf Results.}
We start by considering the ground state energy $E_N$ of a balanced ensemble of $N$ interacting fermions in the trap. In order to appropriately compare ensembles with different numbers of particles, in \mbox{Fig.\  \ref{fig:rescaledEnergies}} we plot the dimensionless quantity $\E_N=E_N/E_N^{(0)}$ as a function of the rescaled interaction strength $\gamma$. The result is analytic for the simplest case of two $(1+1)$ particles, and is the solution of the implicit equation \cite{Busch1998}
\beq
\frac{1}{g}=\frac{\Gamma(3/2-E_2/2)}{\sqrt{2}(E_2-1)\Gamma(1-E_2/2)}.
\eeq
In the thermodynamic limit of an infinite number of particles, instead, the analytic result $E_\infty$ can be obtained by applying the local density approximation (LDA) to the solution of the Gaudin-Yang (GY) integral equations describing a homogeneous gas \cite{Astrakharchik2004}.  We start by noticing that the results in the two extreme limits are actually surprisingly close to each other. As shown in the inset of Fig.\  \ref{fig:rescaledEnergies}, the many-body and two-body rescaled energies differ by less than 0.05 over the complete range of interaction strengths. This poses a serious challenge to the numerical calculation of the energies. Nonetheless, we see that the results of both CC and FCI lie nicely between the two curves, thereby on one side showing their accuracy, and on the other providing a further confirmation of the validity of LDA for computing the energy of a trapped 1D gas.

The size of the computational space diverges exponentially with the number of particles within exact diagonalization, so that with this method we could obtain converged results only for very small systems, containing at most three pairs of atoms ($N=6$), compatibly with what was recently published in Ref.\ \cite{DAmico2015}. On the other hand, only a carefully chosen subset of the total Hilbert space is retained in the CC calculations, so that this method allows us to investigate much larger systems, even up to 40+40 particles. The range of interaction strengths we are able to explore however slowly shrinks with the size of the system $N$, as the complexity of the calculation scales with the bare interaction strength $g$, rather than with its rescaled counterpart $\gamma\propto g/\sqrt{N}$.

\begin{figure}
\centering
\includegraphics[width=\columnwidth]{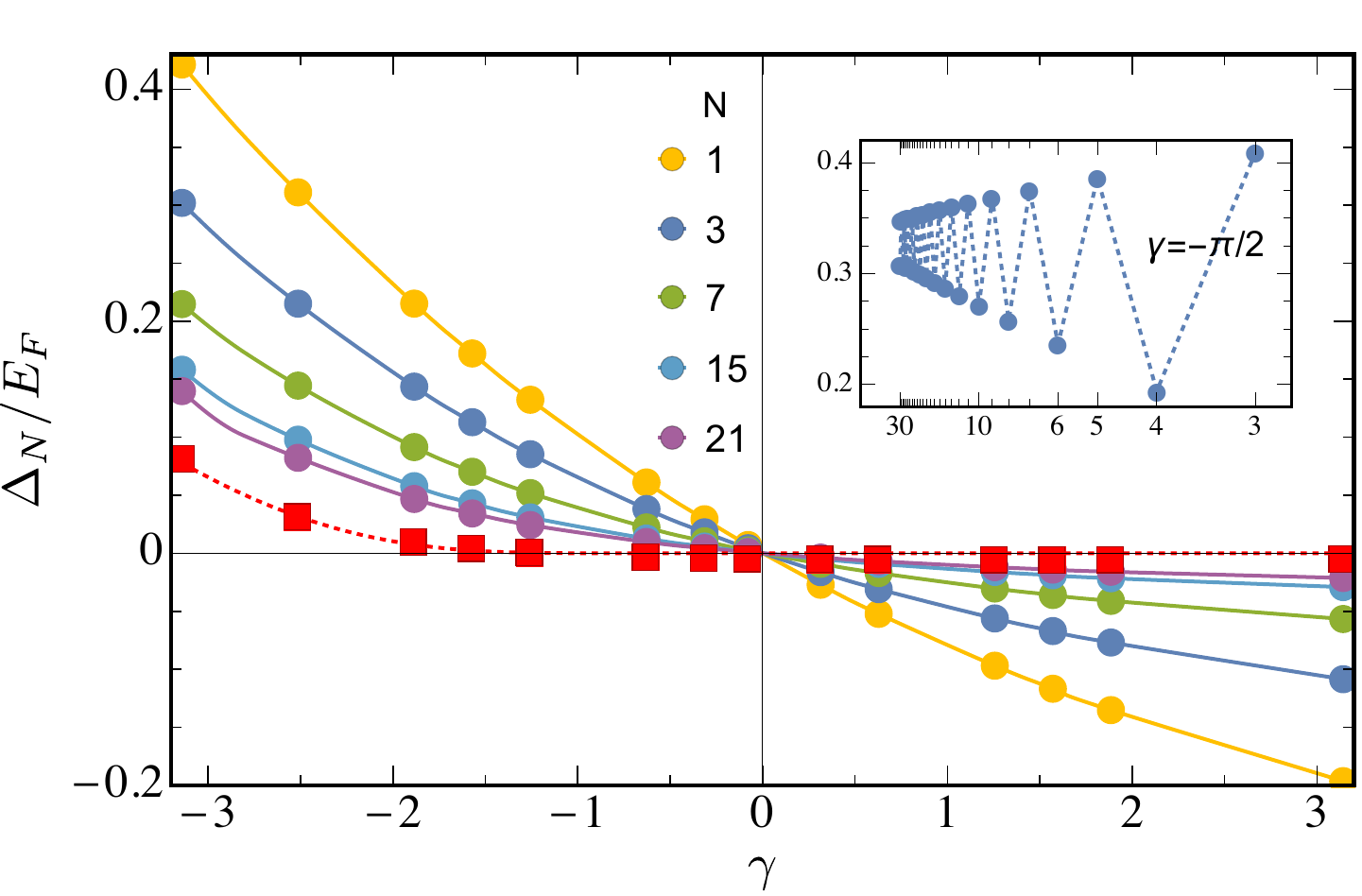}
\caption{BCS pairing gap $\Delta_N$ of a few-fermion system with $N$ particles. Filled circles indicate CC results for increasing $N$, and the red squares are their extrapolation to $N\rightarrow\infty$. The red dotted line is the thermodynamic result, Eq.\ \eqref{BCSpairingGap}. Inset: chemical potential $\mu_N$ [in units of $N\hbar\omega$] vs.\ $1/N$, evaluated at $\gamma=-\pi/2$.}
\label{fig:pairingGap}
\end{figure}


To investigate in detail the continuous crossover from few- to many-body systems, in Fig.\ \ref{fig:energyDifferences} we show our results after subtracting the analytical two-body contribution $\E_2$. Even after the subtraction, the energies are shown to converge remarkably fast to the thermodynamic limit (red dotted line). While our CC and FCI results closely match for 2+2 particles, at this level of precision one notices that the FCI results for 3+3 particles start to deviate from the expected behavior: as an example, for $\gamma\gtrsim2$ the 3+3 FCI results (orange squares) unphysically cross the 2+2 results (blue symbols), even after a very time-consuming calculation (a week of CPU time to get one not-yet-converged FCI point, compared to two hours for a converged CCSD(T) point).
 The insets of Fig. \ref{fig:energyDifferences} show the rescaled CC few-body energies plotted at fixed interaction strengths versus $1/N$. For large $N$ the few-body results smoothly extrapolate to the GY+LDA thermodynamic limit (green dots), while for small $g$ (i.e., small $\gamma$ and small $N$) the results match the first order perturbation result
\beq
E_N=E_N^{(0)}+g\int_{-\infty}^{\infty}{\rm d}z\, n_{0}^2+\mathcal{O}(g^2),
\label{enDiffMeanField}
\eeq
where $n_{0}=\sum_{i=0}^{N/2-1}|\phi_i(z)|^2$ is the density of a gas of $N/2$ identical fermions. 
We note in passing that, in the limit of small $g$ and large $N$, the latter equation recovers the expected weak-coupling LDA result, $E_N=E_N^{(0)}+4gN^{3/2}/(3\pi^2 a_z)+\mathcal{O}(g^2)$~\cite{Astrakharchik2005}.


We turn now to consider the BCS pairing gap, which for this few-body system we define as $\Delta_N=E_N-(E_{N+1}+E_{N-1})/2$, for odd values of $N$ (with $N_\uparrow=N_\downarrow+1$). The results of our CC calculations are shown in Fig.\ \ref{fig:pairingGap}. The BCS pairing gap equals half the spin gap, and in the thermodynamic limit for a homogeneous system it is identically zero for repulsive interactions, while it has a characteristic non-analytic behavior for small attractive interactions. The GY result for a homogeneous gas \cite{Fuchs2004} may be adapted to describe a trapped system by replacing the homogeneous Fermi momentum $\pi n/2$ ($n$ being the total density) with its value at the center of the trap, $k_F\sim\sqrt{N}/a_z$ (for $N\gg1$). 
The specific choice of the momentum at the center of the trap can be justified by the fact that, even if the extra particles are added at the edges of the cloud, their presence will cause an overall reorganization of the gas density profile, so that the resulting energy gap will be sensitive to the typical momentum $k_F$. This procedure yields
\beq
\Delta_\infty=8E_F\sqrt{-\frac{\gamma}{2\pi^3}}\exp\left(\frac{\pi^2}{\gamma}\right)
\label{BCSpairingGap}
\eeq
Our few-body results are, as expected, analytic across the non-interacting point.
However, one can clearly see how the progressive build-up of the Fermi sea gives rise to the expected non-analytic behavior in the weak interaction limit, as predicted by Eq.~\eqref{BCSpairingGap}.
The inset of Fig.~\ref{fig:pairingGap} shows instead the behavior of the chemical potential $\mu_N=E_{N+1}-E_N$, which is characterized by a pronounced even/odd effect due to the shell structure of the trap. At odds with the energy, 
the approach to the thermodynamic behavior is actually very slow for both the pairing gap and chemical potential.


\begin{figure}
\centering
\includegraphics[width=\columnwidth]{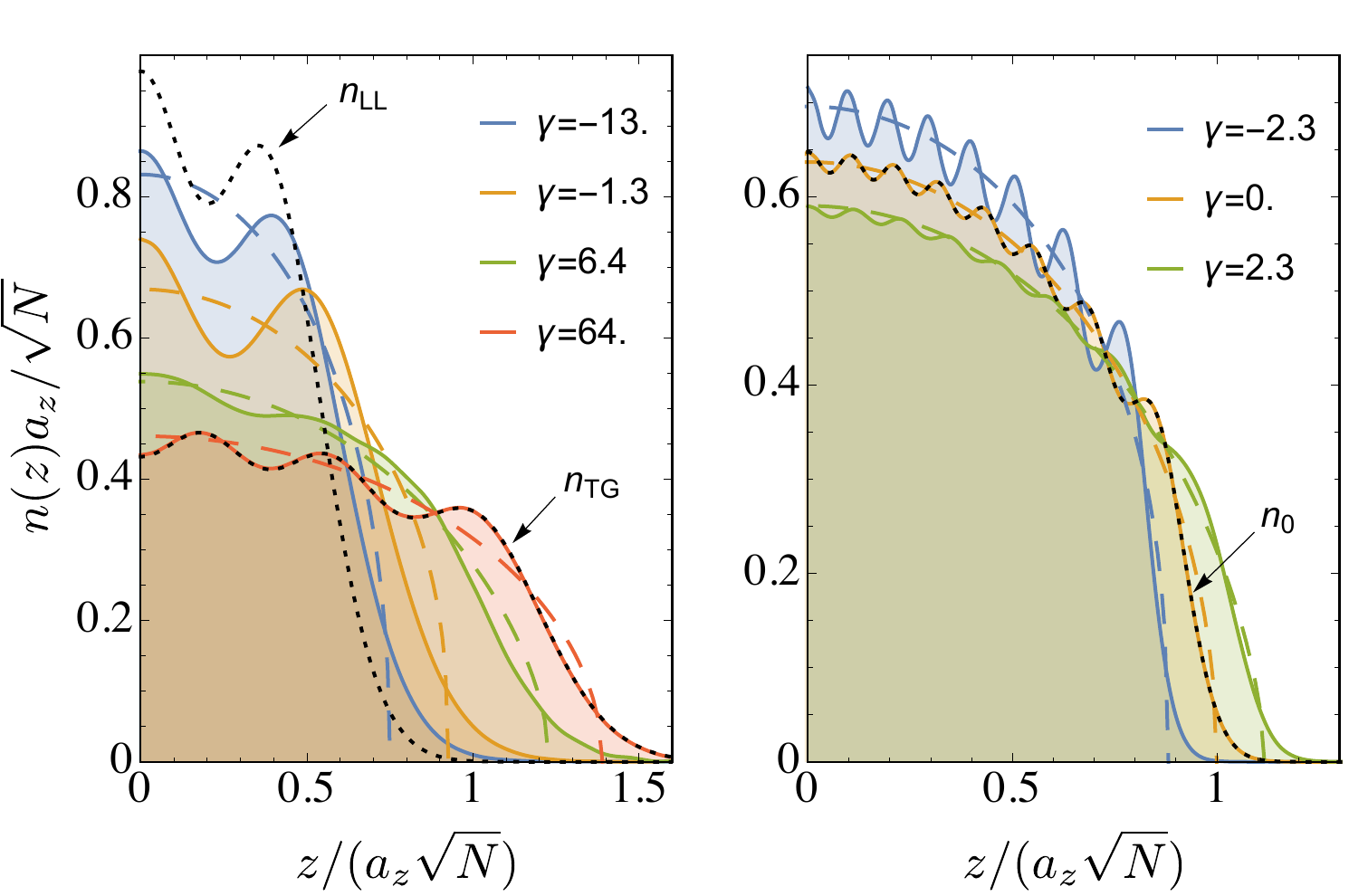}
\caption{Density of a balanced two-component Fermi gas. The solid lines in the left panel display FCI results for 3+3 particles, while the ones in the right panel show CC results for 15+15 atoms. The dashed lines are the GY+LDA results, and the black dotted lines are analytic results for $\gamma\ll-1$ (LL), $\gamma=0$ (0), and $\gamma\gg1$ (TG).}
\label{fig:rescaledDensities}
\end{figure}

Finally, of great interest is the density profile of the trapped cloud. 
The density at a point $z_0$ is obtained as the expectation value of the operator $\sum_{i=1}^{N}\delta(z_i-z_0)$ within the FCI calculations, and in the CC approach from
the Hellmann-Feynman theorem by adding a perturbation of the form $\lambda\phi_i(z_0)\phi_j(z_0)$ to the $ij$-th element of the one-particle Hamiltonian matrix, and taking the derivative of the energy with respect to $\lambda$.
The density profiles we obtain are shown in Fig.\ \ref{fig:rescaledDensities}.
These become broader as the interaction strength grows increasingly more repulsive, and display a regular series of peaks. 
One finds exactly $N/2$ peaks in the LL limit of strong attraction ($\gamma\ll-1$), where pairs of $\up\down$ fermions become tightly-bound bosonic hard-core dimers, and the density approaches $n_{\rm LL}=2\sum_{i=0}^{N/2-1}|\tilde\phi_i|^2$, with $\tilde\phi_i(z)$ the wavefunction of the $i$-th level for a particle of mass $2m$. $N/2$ peaks are present also in the non-interacting case, where two distinguishable fermions occupy the same orbital, and the density becomes $2n_{0}$
. The number of peaks then smoothly evolves to $N$ in the TG limit ($\gamma\gg1$) of ``fermionized" fermions, where due to the strong repulsion even distinguishable fermions must occupy different one-particle levels, and the density approaches $n_{\rm TG}=\sum_{i=0}^{N-1}|\phi_i|^2$. In the thermodynamic limit, the GY+LDA analysis \cite{Astrakharchik2004} predicts a density with the typical Thomas-Fermi (TF) profile of an inverted parabola, $n_{\rm TF}(z)\propto(1-z^2/R_{\rm TF}^2)$, where the radius $R_{\rm TF}$ varies between $\sqrt{N/2}a_z$ in the LL limit and $\sqrt{2N}a_z$ in the TG limit, being exactly equal to $\sqrt{N}a_z$ for an ideal gas.  The two approaches used here prove to be somehow complementary. The FCI method allows us to consider very strong repulsive interactions and enter the extreme ``fermionized" regimes: the hard-core bosonic LL gas, and the fermionized atomic TG gas. Such strong repulsion is out of the reach of present CC calculations, since in this regime both the Hartree-Fock and the correlation energies diverge. However, with CC we are able to address much larger systems and explore beyond mean-field corrections. In particular, this method gives direct access to  non-perturbative effects such as the progressive build up of a non-analytic behavior for the pairing gap.

Summarizing, we presented a detailed analysis of the static properties of a two component Fermi gas in a 1D harmonic trap. 
We computed with high accuracy the energy, the chemical potential, the pairing gap, and density profiles for $N$ ranging from a few to a few tens, and for a broad range of interaction strengths, well beyond the mean-field regime. Our predictions for $\E_N$, $\mu_N$ and $\Delta_N$ may be tested directly by RF spectroscopy or by tunneling measurements, as done in Refs.~\cite{Wenz2013,Zurn2013}. The CC method proved to be extremely well suited to studying harmonically trapped fermions, and we foresee that it will equally well describe more complex potentials, such as double wells or microtrap arrays \cite{Murmann2015}, and dipolar systems.  The method could moreover be generalized to 2D and 3D, or to few strongly-interacting bosonic atoms. These studies would, in particular, be crucial to understand whether the rapid convergence of the system's energy found here is mainly due to the short-range character of the interactions, to the 1D confinement, or to the harmonic external potential.


\begin{acknowledgments}
{\bf Acknowledgments} We wish to thank Grigori Astrakharchik, Miguel-Angel Garcia-March, Bogumi{\l} Jeziorski, Jesper Levinsen, Meera Parish, and Leticia Tarruell for fruitful discussions.
We acknowledge support from the EU grants ERC AdG OSYRIS, FP7 SIQS and EQuaM, FETPROACT QUIC, Spanish Ministry grant FOQUS (FIS2013-46768-P), the Generalitat de Catalunya project 2014 SGR
874, Fundaci\'o Cellex, the Ram\'on y Cajal programme, EU Marie Curie COFUND action (ICFOnest),  FNP START and MISTRZ programs, Polish National Science Centre (ST4/04929), and the PL-Grid Infrastructure.
\end{acknowledgments}

\bibliography{1d_trapped_fermions_QC}

\begin{thebibliography}{66}%
\makeatletter
\providecommand \@ifxundefined [1]{%
 \@ifx{#1\undefined}
}%
\providecommand \@ifnum [1]{%
 \ifnum #1\expandafter \@firstoftwo
 \else \expandafter \@secondoftwo
 \fi
}%
\providecommand \@ifx [1]{%
 \ifx #1\expandafter \@firstoftwo
 \else \expandafter \@secondoftwo
 \fi
}%
\providecommand \natexlab [1]{#1}%
\providecommand \enquote  [1]{``#1''}%
\providecommand \bibnamefont  [1]{#1}%
\providecommand \bibfnamefont [1]{#1}%
\providecommand \citenamefont [1]{#1}%
\providecommand \href@noop [0]{\@secondoftwo}%
\providecommand \href [0]{\begingroup \@sanitize@url \@href}%
\providecommand \@href[1]{\@@startlink{#1}\@@href}%
\providecommand \@@href[1]{\endgroup#1\@@endlink}%
\providecommand \@sanitize@url [0]{\catcode `\\12\catcode `\$12\catcode
  `\&12\catcode `\#12\catcode `\^12\catcode `\_12\catcode `\%12\relax}%
\providecommand \@@startlink[1]{}%
\providecommand \@@endlink[0]{}%
\providecommand \url  [0]{\begingroup\@sanitize@url \@url }%
\providecommand \@url [1]{\endgroup\@href {#1}{\urlprefix }}%
\providecommand \urlprefix  [0]{URL }%
\providecommand \Eprint [0]{\href }%
\providecommand \doibase [0]{http://dx.doi.org/}%
\providecommand \selectlanguage [0]{\@gobble}%
\providecommand \bibinfo  [0]{\@secondoftwo}%
\providecommand \bibfield  [0]{\@secondoftwo}%
\providecommand \translation [1]{[#1]}%
\providecommand \BibitemOpen [0]{}%
\providecommand \bibitemStop [0]{}%
\providecommand \bibitemNoStop [0]{.\EOS\space}%
\providecommand \EOS [0]{\spacefactor3000\relax}%
\providecommand \BibitemShut  [1]{\csname bibitem#1\endcsname}%
\let\auto@bib@innerbib\@empty
\bibitem [{\citenamefont {Lewenstein}\ \emph {et~al.}(2012)\citenamefont
  {Lewenstein}, \citenamefont {Sanpera},\ and\ \citenamefont
  {Ahufinger}}]{Lewenstein2012}%
  \BibitemOpen
  \bibfield  {author} {\bibinfo {author} {\bibfnamefont {M.}~\bibnamefont
  {Lewenstein}}, \bibinfo {author} {\bibfnamefont {A.}~\bibnamefont {Sanpera}},
  \ and\ \bibinfo {author} {\bibfnamefont {V.}~\bibnamefont {Ahufinger}},\
  }\href@noop {} {\emph {\bibinfo {title} {Ultracold atoms in Optical Lattices:
  simulating quantum many body physics}}}\ (\bibinfo  {publisher} {Oxford
  University Press},\ \bibinfo {address} {Oxford},\ \bibinfo {year}
  {2012})\BibitemShut {NoStop}%
\bibitem [{\citenamefont {Bloch}\ \emph {et~al.}(2008)\citenamefont {Bloch},
  \citenamefont {Dalibard},\ and\ \citenamefont {Zwerger}}]{Bloch2008}%
  \BibitemOpen
  \bibfield  {author} {\bibinfo {author} {\bibfnamefont {I.}~\bibnamefont
  {Bloch}}, \bibinfo {author} {\bibfnamefont {J.}~\bibnamefont {Dalibard}}, \
  and\ \bibinfo {author} {\bibfnamefont {W.}~\bibnamefont {Zwerger}},\
  }\bibfield  {title} {\bibinfo {title} {\emph {Many-body physics with
  ultracold gases}},\ }\href {\doibase 10.1103/RevModPhys.80.885} {\bibfield
  {journal} {\bibinfo  {journal} {Rev. Mod. Phys.}\ }\textbf {\bibinfo {volume}
  {80}},\ \bibinfo {pages} {885} (\bibinfo {year} {2008})}\BibitemShut
  {NoStop}%
\bibitem [{\citenamefont {Salomon}\ \emph {et~al.}(2012)\citenamefont
  {Salomon}, \citenamefont {Shlyapnikov},\ and\ \citenamefont
  {Cugliandolo}}]{HouchesVol94}%
  \BibitemOpen
  \bibinfo {editor} {\bibfnamefont {C.}~\bibnamefont {Salomon}}, \bibinfo
  {editor} {\bibfnamefont {G.~V.}\ \bibnamefont {Shlyapnikov}}, \ and\ \bibinfo
  {editor} {\bibfnamefont {L.~F.}\ \bibnamefont {Cugliandolo}},\ eds.,\
  \href@noop {} {\emph {\bibinfo {title} {Many-Body Physics with Ultracold
  Gases}}},\ \bibinfo {series} {Lecture Notes of Les Houches Summer School},
  Vol.~\bibinfo {volume} {94}\ (\bibinfo  {publisher} {Oxford University
  Press},\ \bibinfo {year} {2012})\BibitemShut {NoStop}%
\bibitem [{\citenamefont {Lewenstein}\ \emph {et~al.}(2007)\citenamefont
  {Lewenstein}, \citenamefont {Sanpera}, \citenamefont {Ahufinger},
  \citenamefont {Damski}, \citenamefont {Sen(De)},\ and\ \citenamefont
  {Sen}}]{Lewenstein2007}%
  \BibitemOpen
  \bibfield  {author} {\bibinfo {author} {\bibfnamefont {M.}~\bibnamefont
  {Lewenstein}}, \bibinfo {author} {\bibfnamefont {A.}~\bibnamefont {Sanpera}},
  \bibinfo {author} {\bibfnamefont {V.}~\bibnamefont {Ahufinger}}, \bibinfo
  {author} {\bibfnamefont {B.}~\bibnamefont {Damski}}, \bibinfo {author}
  {\bibfnamefont {A.}~\bibnamefont {Sen(De)}}, \ and\ \bibinfo {author}
  {\bibfnamefont {U.}~\bibnamefont {Sen}},\ }\bibfield  {title} {\bibinfo
  {title} {\emph {Ultracold atomic gases in optical lattices: mimicking
  condensed matter physics and beyond}},\ }\href {\doibase
  10.1080/00018730701223200} {\bibfield  {journal} {\bibinfo  {journal} {Adv.
  Phys.}\ }\textbf {\bibinfo {volume} {56}},\ \bibinfo {pages} {243} (\bibinfo
  {year} {2007})}\BibitemShut {NoStop}%
\bibitem [{\citenamefont {Giamarchi}(2004)}]{Giamarchi}%
  \BibitemOpen
  \bibfield  {author} {\bibinfo {author} {\bibfnamefont {T.}~\bibnamefont
  {Giamarchi}},\ }\href@noop {} {\emph {\bibinfo {title} {Quantum Physics in
  One Dimension}}}\ (\bibinfo  {publisher} {Clarendon Press},\ \bibinfo
  {address} {Oxford},\ \bibinfo {year} {2004})\BibitemShut {NoStop}%
\bibitem [{\citenamefont {Moritz}\ \emph {et~al.}(2003)\citenamefont {Moritz},
  \citenamefont {St\"oferle}, \citenamefont {K\"ohl},\ and\ \citenamefont
  {Esslinger}}]{Tilman1D}%
  \BibitemOpen
  \bibfield  {author} {\bibinfo {author} {\bibfnamefont {H.}~\bibnamefont
  {Moritz}}, \bibinfo {author} {\bibfnamefont {T.}~\bibnamefont {St\"oferle}},
  \bibinfo {author} {\bibfnamefont {M.}~\bibnamefont {K\"ohl}}, \ and\ \bibinfo
  {author} {\bibfnamefont {T.}~\bibnamefont {Esslinger}},\ }\bibfield  {title}
  {\bibinfo {title} {\emph {Exciting Collective Oscillations in a Trapped 1D
  Gas}},\ }\href {\doibase 10.1103/PhysRevLett.91.250402} {\bibfield  {journal}
  {\bibinfo  {journal} {Phys. Rev. Lett.}\ }\textbf {\bibinfo {volume} {91}},\
  \bibinfo {pages} {250402} (\bibinfo {year} {2003})}\BibitemShut {NoStop}%
\bibitem [{\citenamefont {Kinoshita}\ \emph {et~al.}(2004)\citenamefont
  {Kinoshita}, \citenamefont {Wenger},\ and\ \citenamefont {Weiss}}]{Weiss1D}%
  \BibitemOpen
  \bibfield  {author} {\bibinfo {author} {\bibfnamefont {T.}~\bibnamefont
  {Kinoshita}}, \bibinfo {author} {\bibfnamefont {T.}~\bibnamefont {Wenger}}, \
  and\ \bibinfo {author} {\bibfnamefont {D.~S.}\ \bibnamefont {Weiss}},\
  }\bibfield  {title} {\bibinfo {title} {\emph {Observation of a
  One-Dimensional Tonks-Girardeau Gas}},\ }\href {\doibase
  10.1126/science.1100700} {\bibfield  {journal} {\bibinfo  {journal}
  {Science}\ }\textbf {\bibinfo {volume} {305}},\ \bibinfo {pages} {1125}
  (\bibinfo {year} {2004})}\BibitemShut {NoStop}%
\bibitem [{\citenamefont {Paredes}\ \emph {et~al.}(2004)\citenamefont
  {Paredes}, \citenamefont {Widera}, \citenamefont {Murg}, \citenamefont
  {Mandel}, \citenamefont {Foelling}, \citenamefont {Cirac}, \citenamefont
  {Shlyapnikov}, \citenamefont {Hansch},\ and\ \citenamefont {Bloch}}]{Belen}%
  \BibitemOpen
  \bibfield  {author} {\bibinfo {author} {\bibfnamefont {B.}~\bibnamefont
  {Paredes}}, \bibinfo {author} {\bibfnamefont {A.}~\bibnamefont {Widera}},
  \bibinfo {author} {\bibfnamefont {V.}~\bibnamefont {Murg}}, \bibinfo {author}
  {\bibfnamefont {O.}~\bibnamefont {Mandel}}, \bibinfo {author} {\bibfnamefont
  {S.}~\bibnamefont {Foelling}}, \bibinfo {author} {\bibfnamefont
  {I.}~\bibnamefont {Cirac}}, \bibinfo {author} {\bibfnamefont {G.~V.}\
  \bibnamefont {Shlyapnikov}}, \bibinfo {author} {\bibfnamefont {T.~W.}\
  \bibnamefont {Hansch}}, \ and\ \bibinfo {author} {\bibfnamefont
  {I.}~\bibnamefont {Bloch}},\ }\bibfield  {title} {\bibinfo {title} {\emph
  {{Tonks-Girardeau gas of ultracold atoms in an optical lattice}}},\ }\href
  {\doibase 10.1038/nature02530} {\bibfield  {journal} {\bibinfo  {journal}
  {Nature}\ }\textbf {\bibinfo {volume} {429}},\ \bibinfo {pages} {277}
  (\bibinfo {year} {2004})}\BibitemShut {NoStop}%
\bibitem [{\citenamefont {Haller}\ \emph {et~al.}(2009)\citenamefont {Haller},
  \citenamefont {Gustavsson}, \citenamefont {Mark}, \citenamefont {Danzl},
  \citenamefont {Hart}, \citenamefont {Pupillo},\ and\ \citenamefont
  {N\"agerl}}]{Haller2009}%
  \BibitemOpen
  \bibfield  {author} {\bibinfo {author} {\bibfnamefont {E.}~\bibnamefont
  {Haller}}, \bibinfo {author} {\bibfnamefont {M.}~\bibnamefont {Gustavsson}},
  \bibinfo {author} {\bibfnamefont {M.~J.}\ \bibnamefont {Mark}}, \bibinfo
  {author} {\bibfnamefont {J.~G.}\ \bibnamefont {Danzl}}, \bibinfo {author}
  {\bibfnamefont {R.}~\bibnamefont {Hart}}, \bibinfo {author} {\bibfnamefont
  {G.}~\bibnamefont {Pupillo}}, \ and\ \bibinfo {author} {\bibfnamefont
  {H.-C.}\ \bibnamefont {N\"agerl}},\ }\bibfield  {title} {\bibinfo {title}
  {\emph {Realization of an Excited, Strongly Correlated Quantum Gas Phase}},\
  }\href {\doibase 10.1126/science.1175850} {\bibfield  {journal} {\bibinfo
  {journal} {Science}\ }\textbf {\bibinfo {volume} {325}},\ \bibinfo {pages}
  {1224} (\bibinfo {year} {2009})}\BibitemShut {NoStop}%
\bibitem [{\citenamefont {Pagano}\ \emph {et~al.}(2014)\citenamefont {Pagano},
  \citenamefont {Mancini}, \citenamefont {Cappellini}, \citenamefont
  {Lombardi}, \citenamefont {Sch\"afer}, \citenamefont {Hu}, \citenamefont
  {Liu}, \citenamefont {Catani}, \citenamefont {Sias}, \citenamefont
  {Inguscio},\ and\ \citenamefont {Fallani}}]{Pagano2014}%
  \BibitemOpen
  \bibfield  {author} {\bibinfo {author} {\bibfnamefont {G.}~\bibnamefont
  {Pagano}}, \bibinfo {author} {\bibfnamefont {M.}~\bibnamefont {Mancini}},
  \bibinfo {author} {\bibfnamefont {G.}~\bibnamefont {Cappellini}}, \bibinfo
  {author} {\bibfnamefont {P.}~\bibnamefont {Lombardi}}, \bibinfo {author}
  {\bibfnamefont {F.}~\bibnamefont {Sch\"afer}}, \bibinfo {author}
  {\bibfnamefont {H.}~\bibnamefont {Hu}}, \bibinfo {author} {\bibfnamefont
  {X.-J.}\ \bibnamefont {Liu}}, \bibinfo {author} {\bibfnamefont
  {J.}~\bibnamefont {Catani}}, \bibinfo {author} {\bibfnamefont
  {C.}~\bibnamefont {Sias}}, \bibinfo {author} {\bibfnamefont {M.}~\bibnamefont
  {Inguscio}}, \ and\ \bibinfo {author} {\bibfnamefont {L.}~\bibnamefont
  {Fallani}},\ }\bibfield  {title} {\bibinfo {title} {\emph {A one-dimensional
  liquid of fermions with tunable spin}},\ }\href {\doibase 10.1038/nphys2878}
  {\bibfield  {journal} {\bibinfo  {journal} {Nature Phys.}\ }\textbf {\bibinfo
  {volume} {10}},\ \bibinfo {pages} {198} (\bibinfo {year} {2014})}\BibitemShut
  {NoStop}%
\bibitem [{\citenamefont {Serwane}\ \emph {et~al.}(2011)\citenamefont
  {Serwane}, \citenamefont {Z\"urn}, \citenamefont {Lompe}, \citenamefont
  {Ottenstein}, \citenamefont {Wenz},\ and\ \citenamefont
  {Jochim}}]{Serwane2011}%
  \BibitemOpen
  \bibfield  {author} {\bibinfo {author} {\bibfnamefont {F.}~\bibnamefont
  {Serwane}}, \bibinfo {author} {\bibfnamefont {G.}~\bibnamefont {Z\"urn}},
  \bibinfo {author} {\bibfnamefont {T.}~\bibnamefont {Lompe}}, \bibinfo
  {author} {\bibfnamefont {T.~B.}\ \bibnamefont {Ottenstein}}, \bibinfo
  {author} {\bibfnamefont {A.~N.}\ \bibnamefont {Wenz}}, \ and\ \bibinfo
  {author} {\bibfnamefont {S.}~\bibnamefont {Jochim}},\ }\bibfield  {title}
  {\bibinfo {title} {\emph {Deterministic Preparation of a Tunable Few-Fermion
  System}},\ }\href {\doibase 10.1126/science.1201351} {\bibfield  {journal}
  {\bibinfo  {journal} {Science}\ }\textbf {\bibinfo {volume} {332}},\ \bibinfo
  {pages} {336} (\bibinfo {year} {2011})}\BibitemShut {NoStop}%
\bibitem [{\citenamefont {Z\"urn}\ \emph {et~al.}(2012)\citenamefont {Z\"urn},
  \citenamefont {Serwane}, \citenamefont {Lompe}, \citenamefont {Wenz},
  \citenamefont {Ries}, \citenamefont {Bohn},\ and\ \citenamefont
  {Jochim}}]{Zurn2012}%
  \BibitemOpen
  \bibfield  {author} {\bibinfo {author} {\bibfnamefont {G.}~\bibnamefont
  {Z\"urn}}, \bibinfo {author} {\bibfnamefont {F.}~\bibnamefont {Serwane}},
  \bibinfo {author} {\bibfnamefont {T.}~\bibnamefont {Lompe}}, \bibinfo
  {author} {\bibfnamefont {A.~N.}\ \bibnamefont {Wenz}}, \bibinfo {author}
  {\bibfnamefont {M.~G.}\ \bibnamefont {Ries}}, \bibinfo {author}
  {\bibfnamefont {J.~E.}\ \bibnamefont {Bohn}}, \ and\ \bibinfo {author}
  {\bibfnamefont {S.}~\bibnamefont {Jochim}},\ }\bibfield  {title} {\bibinfo
  {title} {\emph {Fermionization of Two Distinguishable Fermions}},\ }\href
  {\doibase 10.1103/PhysRevLett.108.075303} {\bibfield  {journal} {\bibinfo
  {journal} {Phys. Rev. Lett.}\ }\textbf {\bibinfo {volume} {108}},\ \bibinfo
  {pages} {075303} (\bibinfo {year} {2012})}\BibitemShut {NoStop}%
\bibitem [{\citenamefont {Wenz}\ \emph {et~al.}(2013)\citenamefont {Wenz},
  \citenamefont {Z{\"u}rn}, \citenamefont {Murmann}, \citenamefont {Brouzos},
  \citenamefont {Lompe},\ and\ \citenamefont {Jochim}}]{Wenz2013}%
  \BibitemOpen
  \bibfield  {author} {\bibinfo {author} {\bibfnamefont {A.}~\bibnamefont
  {Wenz}}, \bibinfo {author} {\bibfnamefont {G.}~\bibnamefont {Z{\"u}rn}},
  \bibinfo {author} {\bibfnamefont {S.}~\bibnamefont {Murmann}}, \bibinfo
  {author} {\bibfnamefont {I.}~\bibnamefont {Brouzos}}, \bibinfo {author}
  {\bibfnamefont {T.}~\bibnamefont {Lompe}}, \ and\ \bibinfo {author}
  {\bibfnamefont {S.}~\bibnamefont {Jochim}},\ }\bibfield  {title} {\bibinfo
  {title} {\emph {From few to many: observing the formation of a Fermi sea one
  atom at a time}},\ }\href {\doibase 10.1126/science.1240516} {\bibfield
  {journal} {\bibinfo  {journal} {Science}\ }\textbf {\bibinfo {volume}
  {342}},\ \bibinfo {pages} {457} (\bibinfo {year} {2013})}\BibitemShut
  {NoStop}%
\bibitem [{\citenamefont {Z\"urn}\ \emph {et~al.}(2013)\citenamefont {Z\"urn},
  \citenamefont {Wenz}, \citenamefont {Murmann}, \citenamefont {Bergschneider},
  \citenamefont {Lompe},\ and\ \citenamefont {Jochim}}]{Zurn2013}%
  \BibitemOpen
  \bibfield  {author} {\bibinfo {author} {\bibfnamefont {G.}~\bibnamefont
  {Z\"urn}}, \bibinfo {author} {\bibfnamefont {A.~N.}\ \bibnamefont {Wenz}},
  \bibinfo {author} {\bibfnamefont {S.}~\bibnamefont {Murmann}}, \bibinfo
  {author} {\bibfnamefont {A.}~\bibnamefont {Bergschneider}}, \bibinfo {author}
  {\bibfnamefont {T.}~\bibnamefont {Lompe}}, \ and\ \bibinfo {author}
  {\bibfnamefont {S.}~\bibnamefont {Jochim}},\ }\bibfield  {title} {\bibinfo
  {title} {\emph {Pairing in Few-Fermion Systems with Attractive
  Interactions}},\ }\href {\doibase 10.1103/PhysRevLett.111.175302} {\bibfield
  {journal} {\bibinfo  {journal} {Phys. Rev. Lett.}\ }\textbf {\bibinfo
  {volume} {111}},\ \bibinfo {pages} {175302} (\bibinfo {year}
  {2013})}\BibitemShut {NoStop}%
\bibitem [{\citenamefont {Murmann}\ \emph {et~al.}(2015)\citenamefont
  {Murmann}, \citenamefont {Bergschneider}, \citenamefont {Klinkhamer},
  \citenamefont {Z\"urn}, \citenamefont {Lompe},\ and\ \citenamefont
  {Jochim}}]{Murmann2015}%
  \BibitemOpen
  \bibfield  {author} {\bibinfo {author} {\bibfnamefont {S.}~\bibnamefont
  {Murmann}}, \bibinfo {author} {\bibfnamefont {A.}~\bibnamefont
  {Bergschneider}}, \bibinfo {author} {\bibfnamefont {V.~M.}\ \bibnamefont
  {Klinkhamer}}, \bibinfo {author} {\bibfnamefont {G.}~\bibnamefont {Z\"urn}},
  \bibinfo {author} {\bibfnamefont {T.}~\bibnamefont {Lompe}}, \ and\ \bibinfo
  {author} {\bibfnamefont {S.}~\bibnamefont {Jochim}},\ }\bibfield  {title}
  {\bibinfo {title} {\emph {Two Fermions in a Double Well: Exploring a
  Fundamental Building Block of the Hubbard Model}},\ }\href {\doibase
  10.1103/PhysRevLett.114.080402} {\bibfield  {journal} {\bibinfo  {journal}
  {Phys. Rev. Lett.}\ }\textbf {\bibinfo {volume} {114}},\ \bibinfo {pages}
  {080402} (\bibinfo {year} {2015})}\BibitemShut {NoStop}%
\bibitem [{\citenamefont {Recati}\ \emph {et~al.}(2003)\citenamefont {Recati},
  \citenamefont {Fedichev}, \citenamefont {Zwerger},\ and\ \citenamefont
  {Zoller}}]{Recati2003}%
  \BibitemOpen
  \bibfield  {author} {\bibinfo {author} {\bibfnamefont {A.}~\bibnamefont
  {Recati}}, \bibinfo {author} {\bibfnamefont {P.~O.}\ \bibnamefont
  {Fedichev}}, \bibinfo {author} {\bibfnamefont {W.}~\bibnamefont {Zwerger}}, \
  and\ \bibinfo {author} {\bibfnamefont {P.}~\bibnamefont {Zoller}},\
  }\bibfield  {title} {\bibinfo {title} {\emph {Spin-Charge Separation in
  Ultracold Quantum Gases}},\ }\href {\doibase 10.1103/PhysRevLett.90.020401}
  {\bibfield  {journal} {\bibinfo  {journal} {Phys. Rev. Lett.}\ }\textbf
  {\bibinfo {volume} {90}},\ \bibinfo {pages} {020401} (\bibinfo {year}
  {2003})}\BibitemShut {NoStop}%
\bibitem [{\citenamefont {Juillet}\ \emph {et~al.}(2004)\citenamefont
  {Juillet}, \citenamefont {Gulminelli},\ and\ \citenamefont
  {Chomaz}}]{Juillet2004}%
  \BibitemOpen
  \bibfield  {author} {\bibinfo {author} {\bibfnamefont {O.}~\bibnamefont
  {Juillet}}, \bibinfo {author} {\bibfnamefont {F.}~\bibnamefont {Gulminelli}},
  \ and\ \bibinfo {author} {\bibfnamefont {P.}~\bibnamefont {Chomaz}},\
  }\bibfield  {title} {\bibinfo {title} {\emph {Exact Pairing Correlations for
  One-Dimensionally Trapped Fermions with Stochastic Mean-Field Wave
  Functions}},\ }\href {\doibase 10.1103/PhysRevLett.92.160401} {\bibfield
  {journal} {\bibinfo  {journal} {Phys. Rev. Lett.}\ }\textbf {\bibinfo
  {volume} {92}},\ \bibinfo {pages} {160401} (\bibinfo {year}
  {2004})}\BibitemShut {NoStop}%
\bibitem [{\citenamefont {Astrakharchik}\ \emph {et~al.}(2004)\citenamefont
  {Astrakharchik}, \citenamefont {Blume}, \citenamefont {Giorgini},\ and\
  \citenamefont {Pitaevskii}}]{Astrakharchik2004}%
  \BibitemOpen
  \bibfield  {author} {\bibinfo {author} {\bibfnamefont {G.~E.}\ \bibnamefont
  {Astrakharchik}}, \bibinfo {author} {\bibfnamefont {D.}~\bibnamefont
  {Blume}}, \bibinfo {author} {\bibfnamefont {S.}~\bibnamefont {Giorgini}}, \
  and\ \bibinfo {author} {\bibfnamefont {L.~P.}\ \bibnamefont {Pitaevskii}},\
  }\bibfield  {title} {\bibinfo {title} {\emph {Interacting Fermions in Highly
  Elongated Harmonic Traps}},\ }\href {\doibase 10.1103/PhysRevLett.93.050402}
  {\bibfield  {journal} {\bibinfo  {journal} {Phys. Rev. Lett.}\ }\textbf
  {\bibinfo {volume} {93}},\ \bibinfo {pages} {050402} (\bibinfo {year}
  {2004})}\BibitemShut {NoStop}%
\bibitem [{\citenamefont {Tokatly}(2004)}]{Tokatly2004}%
  \BibitemOpen
  \bibfield  {author} {\bibinfo {author} {\bibfnamefont {I.~V.}\ \bibnamefont
  {Tokatly}},\ }\bibfield  {title} {\bibinfo {title} {\emph {Dilute Fermi Gas
  in Quasi-One-Dimensional Traps: From Weakly Interacting Fermions via Hard
  Core Bosons to a Weakly Interacting Bose Gas}},\ }\href {\doibase
  10.1103/PhysRevLett.93.090405} {\bibfield  {journal} {\bibinfo  {journal}
  {Phys. Rev. Lett.}\ }\textbf {\bibinfo {volume} {93}},\ \bibinfo {pages}
  {090405} (\bibinfo {year} {2004})}\BibitemShut {NoStop}%
\bibitem [{\citenamefont {Fuchs}\ \emph {et~al.}(2004)\citenamefont {Fuchs},
  \citenamefont {Recati},\ and\ \citenamefont {Zwerger}}]{Fuchs2004}%
  \BibitemOpen
  \bibfield  {author} {\bibinfo {author} {\bibfnamefont {J.~N.}\ \bibnamefont
  {Fuchs}}, \bibinfo {author} {\bibfnamefont {A.}~\bibnamefont {Recati}}, \
  and\ \bibinfo {author} {\bibfnamefont {W.}~\bibnamefont {Zwerger}},\
  }\bibfield  {title} {\bibinfo {title} {\emph {Exactly Solvable Model of the
  BCS-BEC Crossover}},\ }\href {\doibase 10.1103/PhysRevLett.93.090408}
  {\bibfield  {journal} {\bibinfo  {journal} {Phys. Rev. Lett.}\ }\textbf
  {\bibinfo {volume} {93}},\ \bibinfo {pages} {090408} (\bibinfo {year}
  {2004})}\BibitemShut {NoStop}%
\bibitem [{\citenamefont {Carusotto}\ and\ \citenamefont
  {Castin}(2004)}]{Carusotto2004}%
  \BibitemOpen
  \bibfield  {author} {\bibinfo {author} {\bibfnamefont {I.}~\bibnamefont
  {Carusotto}}\ and\ \bibinfo {author} {\bibfnamefont {Y.}~\bibnamefont
  {Castin}},\ }\bibfield  {title} {\bibinfo {title} {\emph {Coherence and
  correlation properties of a one-dimensional attractive Fermi gas}},\ }\href
  {\doibase 10.1016/j.optcom.2004.04.062} {\bibfield  {journal} {\bibinfo
  {journal} {Opt Commun}\ }\textbf {\bibinfo {volume} {243}},\ \bibinfo {pages}
  {81 } (\bibinfo {year} {2004})}\BibitemShut {NoStop}%
\bibitem [{\citenamefont {Astrakharchik}(2005)}]{Astrakharchik2005}%
  \BibitemOpen
  \bibfield  {author} {\bibinfo {author} {\bibfnamefont {G.~E.}\ \bibnamefont
  {Astrakharchik}},\ }\bibfield  {title} {\bibinfo {title} {\emph {Local
  density approximation for a perturbative equation of state}},\ }\href
  {\doibase 10.1103/PhysRevA.72.063620} {\bibfield  {journal} {\bibinfo
  {journal} {Phys. Rev. A}\ }\textbf {\bibinfo {volume} {72}},\ \bibinfo
  {pages} {063620} (\bibinfo {year} {2005})}\BibitemShut {NoStop}%
\bibitem [{\citenamefont {Orso}(2007)}]{Orso2007}%
  \BibitemOpen
  \bibfield  {author} {\bibinfo {author} {\bibfnamefont {G.}~\bibnamefont
  {Orso}},\ }\bibfield  {title} {\bibinfo {title} {\emph {Attractive Fermi
  Gases with Unequal Spin Populations in Highly Elongated Traps}},\ }\href
  {\doibase 10.1103/PhysRevLett.98.070402} {\bibfield  {journal} {\bibinfo
  {journal} {Phys. Rev. Lett.}\ }\textbf {\bibinfo {volume} {98}},\ \bibinfo
  {pages} {070402} (\bibinfo {year} {2007})}\BibitemShut {NoStop}%
\bibitem [{\citenamefont {Hu}\ \emph {et~al.}(2007)\citenamefont {Hu},
  \citenamefont {Liu},\ and\ \citenamefont {Drummond}}]{Hu2007}%
  \BibitemOpen
  \bibfield  {author} {\bibinfo {author} {\bibfnamefont {H.}~\bibnamefont
  {Hu}}, \bibinfo {author} {\bibfnamefont {X.-J.}\ \bibnamefont {Liu}}, \ and\
  \bibinfo {author} {\bibfnamefont {P.~D.}\ \bibnamefont {Drummond}},\
  }\bibfield  {title} {\bibinfo {title} {\emph {Phase Diagram of a Strongly
  Interacting Polarized Fermi Gas in One Dimension}},\ }\href {\doibase
  10.1103/PhysRevLett.98.070403} {\bibfield  {journal} {\bibinfo  {journal}
  {Phys. Rev. Lett.}\ }\textbf {\bibinfo {volume} {98}},\ \bibinfo {pages}
  {070403} (\bibinfo {year} {2007})}\BibitemShut {NoStop}%
\bibitem [{\citenamefont {Colom\'e-Tatch\'e}(2008)}]{ColomeTatche2008}%
  \BibitemOpen
  \bibfield  {author} {\bibinfo {author} {\bibfnamefont {M.}~\bibnamefont
  {Colom\'e-Tatch\'e}},\ }\bibfield  {title} {\bibinfo {title} {\emph
  {Two-component repulsive Fermi gases with population imbalance in elongated
  harmonic traps}},\ }\href {\doibase 10.1103/PhysRevA.78.033612} {\bibfield
  {journal} {\bibinfo  {journal} {Phys. Rev. A}\ }\textbf {\bibinfo {volume}
  {78}},\ \bibinfo {pages} {033612} (\bibinfo {year} {2008})}\BibitemShut
  {NoStop}%
\bibitem [{\citenamefont {Casula}\ \emph {et~al.}(2008)\citenamefont {Casula},
  \citenamefont {Ceperley},\ and\ \citenamefont {Mueller}}]{Casula2008}%
  \BibitemOpen
  \bibfield  {author} {\bibinfo {author} {\bibfnamefont {M.}~\bibnamefont
  {Casula}}, \bibinfo {author} {\bibfnamefont {D.~M.}\ \bibnamefont
  {Ceperley}}, \ and\ \bibinfo {author} {\bibfnamefont {E.~J.}\ \bibnamefont
  {Mueller}},\ }\bibfield  {title} {\bibinfo {title} {\emph {{Quantum Monte
  Carlo study of one-dimensional trapped fermions with attractive contact
  interactions}}},\ }\href {\doibase 10.1103/PhysRevA.78.033607} {\bibfield
  {journal} {\bibinfo  {journal} {Phys. Rev. A}\ }\textbf {\bibinfo {volume}
  {78}},\ \bibinfo {pages} {033607} (\bibinfo {year} {2008})}\BibitemShut
  {NoStop}%
\bibitem [{\citenamefont {S\"offing}\ \emph {et~al.}(2011)\citenamefont
  {S\"offing}, \citenamefont {Bortz},\ and\ \citenamefont
  {Eggert}}]{Soeffing2011}%
  \BibitemOpen
  \bibfield  {author} {\bibinfo {author} {\bibfnamefont {S.~A.}\ \bibnamefont
  {S\"offing}}, \bibinfo {author} {\bibfnamefont {M.}~\bibnamefont {Bortz}}, \
  and\ \bibinfo {author} {\bibfnamefont {S.}~\bibnamefont {Eggert}},\
  }\bibfield  {title} {\bibinfo {title} {\emph {Density profile of interacting
  fermions in a one-dimensional optical trap}},\ }\href {\doibase
  10.1103/PhysRevA.84.021602} {\bibfield  {journal} {\bibinfo  {journal} {Phys.
  Rev. A}\ }\textbf {\bibinfo {volume} {84}},\ \bibinfo {pages} {021602}
  (\bibinfo {year} {2011})}\BibitemShut {NoStop}%
\bibitem [{\citenamefont {Gharashi}\ and\ \citenamefont
  {Blume}(2013)}]{Gharashi2013}%
  \BibitemOpen
  \bibfield  {author} {\bibinfo {author} {\bibfnamefont {S.~E.}\ \bibnamefont
  {Gharashi}}\ and\ \bibinfo {author} {\bibfnamefont {D.}~\bibnamefont
  {Blume}},\ }\bibfield  {title} {\bibinfo {title} {\emph {Correlations of the
  Upper Branch of 1D Harmonically Trapped Two-Component Fermi Gases}},\ }\href
  {\doibase 10.1103/PhysRevLett.111.045302} {\bibfield  {journal} {\bibinfo
  {journal} {Phys. Rev. Lett.}\ }\textbf {\bibinfo {volume} {111}},\ \bibinfo
  {pages} {045302} (\bibinfo {year} {2013})}\BibitemShut {NoStop}%
\bibitem [{\citenamefont {Sowi\'nski}\ \emph {et~al.}(2013)\citenamefont
  {Sowi\'nski}, \citenamefont {Grass}, \citenamefont {Dutta},\ and\
  \citenamefont {Lewenstein}}]{Sowinski2013}%
  \BibitemOpen
  \bibfield  {author} {\bibinfo {author} {\bibfnamefont {T.}~\bibnamefont
  {Sowi\'nski}}, \bibinfo {author} {\bibfnamefont {T.}~\bibnamefont {Grass}},
  \bibinfo {author} {\bibfnamefont {O.}~\bibnamefont {Dutta}}, \ and\ \bibinfo
  {author} {\bibfnamefont {M.}~\bibnamefont {Lewenstein}},\ }\bibfield  {title}
  {\bibinfo {title} {\emph {{Few interacting fermions in a one-dimensional
  harmonic trap}}},\ }\href {\doibase 10.1103/PhysRevA.88.033607} {\bibfield
  {journal} {\bibinfo  {journal} {Phys. Rev. A}\ }\textbf {\bibinfo {volume}
  {88}},\ \bibinfo {pages} {033607} (\bibinfo {year} {2013})}\BibitemShut
  {NoStop}%
\bibitem [{\citenamefont {Astrakharchik}\ and\ \citenamefont
  {Brouzos}(2013)}]{Astrakharchik2013}%
  \BibitemOpen
  \bibfield  {author} {\bibinfo {author} {\bibfnamefont {G.~E.}\ \bibnamefont
  {Astrakharchik}}\ and\ \bibinfo {author} {\bibfnamefont {I.}~\bibnamefont
  {Brouzos}},\ }\bibfield  {title} {\bibinfo {title} {\emph {Trapped
  one-dimensional ideal Fermi gas with a single impurity}},\ }\href {\doibase
  10.1103/PhysRevA.88.021602} {\bibfield  {journal} {\bibinfo  {journal} {Phys.
  Rev. A}\ }\textbf {\bibinfo {volume} {88}},\ \bibinfo {pages} {021602}
  (\bibinfo {year} {2013})}\BibitemShut {NoStop}%
\bibitem [{\citenamefont {Deuretzbacher}\ \emph {et~al.}(2014)\citenamefont
  {Deuretzbacher}, \citenamefont {Becker}, \citenamefont {Bjerlin},
  \citenamefont {Reimann},\ and\ \citenamefont {Santos}}]{Deuretzbacher2014}%
  \BibitemOpen
  \bibfield  {author} {\bibinfo {author} {\bibfnamefont {F.}~\bibnamefont
  {Deuretzbacher}}, \bibinfo {author} {\bibfnamefont {D.}~\bibnamefont
  {Becker}}, \bibinfo {author} {\bibfnamefont {J.}~\bibnamefont {Bjerlin}},
  \bibinfo {author} {\bibfnamefont {S.~M.}\ \bibnamefont {Reimann}}, \ and\
  \bibinfo {author} {\bibfnamefont {L.}~\bibnamefont {Santos}},\ }\bibfield
  {title} {\bibinfo {title} {\emph {Quantum magnetism without lattices in
  strongly interacting one-dimensional spinor gases}},\ }\href {\doibase
  10.1103/PhysRevA.90.013611} {\bibfield  {journal} {\bibinfo  {journal} {Phys.
  Rev. A}\ }\textbf {\bibinfo {volume} {90}},\ \bibinfo {pages} {013611}
  (\bibinfo {year} {2014})}\BibitemShut {NoStop}%
\bibitem [{\citenamefont {{Volosniev}}\ \emph {et~al.}(2014)\citenamefont
  {{Volosniev}}, \citenamefont {{Fedorov}}, \citenamefont {{Jensen}},
  \citenamefont {{Valiente}},\ and\ \citenamefont {{Zinner}}}]{Volosniev2014}%
  \BibitemOpen
  \bibfield  {author} {\bibinfo {author} {\bibfnamefont {A.~G.}\ \bibnamefont
  {{Volosniev}}}, \bibinfo {author} {\bibfnamefont {D.~V.}\ \bibnamefont
  {{Fedorov}}}, \bibinfo {author} {\bibfnamefont {A.~S.}\ \bibnamefont
  {{Jensen}}}, \bibinfo {author} {\bibfnamefont {M.}~\bibnamefont
  {{Valiente}}}, \ and\ \bibinfo {author} {\bibfnamefont {N.~T.}\ \bibnamefont
  {{Zinner}}},\ }\bibfield  {title} {\bibinfo {title} {\emph {Strongly
  interacting confined quantum systems in one dimension}},\ }\href {\doibase
  10.1038/ncomms6300} {\bibfield  {journal} {\bibinfo  {journal}
  {Nat.~Commun.}\ }\textbf {\bibinfo {volume} {5}},\ \bibinfo {pages} {5300}
  (\bibinfo {year} {2014})}\BibitemShut {NoStop}%
\bibitem [{\citenamefont {Lindgren}\ \emph {et~al.}(2014)\citenamefont
  {Lindgren}, \citenamefont {Rotureau}, \citenamefont {Forss\'{e}n},
  \citenamefont {Volosniev},\ and\ \citenamefont {Zinner}}]{Lindgren2014}%
  \BibitemOpen
  \bibfield  {author} {\bibinfo {author} {\bibfnamefont {E.~J.}\ \bibnamefont
  {Lindgren}}, \bibinfo {author} {\bibfnamefont {J.}~\bibnamefont {Rotureau}},
  \bibinfo {author} {\bibfnamefont {C.}~\bibnamefont {Forss\'{e}n}}, \bibinfo
  {author} {\bibfnamefont {A.~G.}\ \bibnamefont {Volosniev}}, \ and\ \bibinfo
  {author} {\bibfnamefont {N.~T.}\ \bibnamefont {Zinner}},\ }\bibfield  {title}
  {\bibinfo {title} {\emph {Fermionization of two-component few-fermion systems
  in a one-dimensional harmonic trap}},\ }\href {\doibase
  10.1088/1367-2630/16/6/063003} {\bibfield  {journal} {\bibinfo  {journal}
  {New J. Phys.}\ }\textbf {\bibinfo {volume} {16}},\ \bibinfo {pages} {063003}
  (\bibinfo {year} {2014})}\BibitemShut {NoStop}%
\bibitem [{\citenamefont {{Levinsen}}\ \emph {et~al.}(2015)\citenamefont
  {{Levinsen}}, \citenamefont {{Massignan}}, \citenamefont {{Bruun}},\ and\
  \citenamefont {{Parish}}}]{Levinsen2015}%
  \BibitemOpen
  \bibfield  {author} {\bibinfo {author} {\bibfnamefont {J.}~\bibnamefont
  {{Levinsen}}}, \bibinfo {author} {\bibfnamefont {P.}~\bibnamefont
  {{Massignan}}}, \bibinfo {author} {\bibfnamefont {G.~M.}\ \bibnamefont
  {{Bruun}}}, \ and\ \bibinfo {author} {\bibfnamefont {M.~M.}\ \bibnamefont
  {{Parish}}},\ }\bibfield  {title} {\bibinfo {title} {\emph {Strong-coupling
  ansatz for the one-dimensional Fermi gas in a harmonic potential}},\ }\href
  {\doibase 10.1126/sciadv.1500197} {\bibfield  {journal} {\bibinfo  {journal}
  {Sci. Adv.}\ }\textbf {\bibinfo {volume} {1}},\ \bibinfo {pages} {e1500197}
  (\bibinfo {year} {2015})}\BibitemShut {NoStop}%
\bibitem [{\citenamefont {D'Amico}\ and\ \citenamefont
  {Rontani}(2015)}]{DAmico2015}%
  \BibitemOpen
  \bibfield  {author} {\bibinfo {author} {\bibfnamefont {P.}~\bibnamefont
  {D'Amico}}\ and\ \bibinfo {author} {\bibfnamefont {M.}~\bibnamefont
  {Rontani}},\ }\bibfield  {title} {\bibinfo {title} {\emph {Pairing of a few
  Fermi atoms in one dimension}},\ }\href {\doibase 10.1103/PhysRevA.91.043610}
  {\bibfield  {journal} {\bibinfo  {journal} {Phys. Rev. A}\ }\textbf {\bibinfo
  {volume} {91}},\ \bibinfo {pages} {043610} (\bibinfo {year}
  {2015})}\BibitemShut {NoStop}%
\bibitem [{\citenamefont {Berger}\ \emph {et~al.}(2015)\citenamefont {Berger},
  \citenamefont {Anderson},\ and\ \citenamefont {Drut}}]{Berger2015}%
  \BibitemOpen
  \bibfield  {author} {\bibinfo {author} {\bibfnamefont {C.~E.}\ \bibnamefont
  {Berger}}, \bibinfo {author} {\bibfnamefont {E.~R.}\ \bibnamefont
  {Anderson}}, \ and\ \bibinfo {author} {\bibfnamefont {J.~E.}\ \bibnamefont
  {Drut}},\ }\bibfield  {title} {\bibinfo {title} {\emph {Energy, contact, and
  density profiles of one-dimensional fermions in a harmonic trap via
  nonuniform-lattice Monte Carlo calculations}},\ }\href {\doibase
  10.1103/PhysRevA.91.053618} {\bibfield  {journal} {\bibinfo  {journal} {Phys.
  Rev. A}\ }\textbf {\bibinfo {volume} {91}},\ \bibinfo {pages} {053618}
  (\bibinfo {year} {2015})}\BibitemShut {NoStop}%
\bibitem [{\citenamefont {Sowi\'nski}\ \emph {et~al.}(2015)\citenamefont
  {Sowi\'nski}, \citenamefont {Gajda},\ and\ \citenamefont
  {Rza{\.z}ewski}}]{Sowinski2015}%
  \BibitemOpen
  \bibfield  {author} {\bibinfo {author} {\bibfnamefont {T.}~\bibnamefont
  {Sowi\'nski}}, \bibinfo {author} {\bibfnamefont {M.}~\bibnamefont {Gajda}}, \
  and\ \bibinfo {author} {\bibfnamefont {K.}~\bibnamefont {Rza{\.z}ewski}},\
  }\bibfield  {title} {\bibinfo {title} {\emph {Pairing in a system of a few
  attractive fermions in a harmonic trap}},\ }\href
  {http://stacks.iop.org/0295-5075/109/i=2/a=26005} {\bibfield  {journal}
  {\bibinfo  {journal} {Europhys. Lett.}\ }\textbf {\bibinfo {volume} {109}},\
  \bibinfo {pages} {26005} (\bibinfo {year} {2015})}\BibitemShut {NoStop}%
\bibitem [{\citenamefont {Dreizler}\ and\ \citenamefont
  {Gross}(2012)}]{Dreizler2012}%
  \BibitemOpen
  \bibfield  {author} {\bibinfo {author} {\bibfnamefont {R.~M.}\ \bibnamefont
  {Dreizler}}\ and\ \bibinfo {author} {\bibfnamefont {E.~K.}\ \bibnamefont
  {Gross}},\ }\href@noop {} {\emph {\bibinfo {title} {Density Functional.
  Theory: An Approach to the. Quantum Many-Body Problem}}}\ (\bibinfo
  {publisher} {Springer},\ \bibinfo {address} {Heidelberg},\ \bibinfo {year}
  {2012})\BibitemShut {NoStop}%
\bibitem [{\citenamefont {Sandvik}(2010)}]{Sandvik2010}%
  \BibitemOpen
  \bibfield  {author} {\bibinfo {author} {\bibfnamefont {A.~W.}\ \bibnamefont
  {Sandvik}},\ }\bibfield  {title} {\bibinfo {title} {\emph {Computational
  Studies of Quantum Spin Systems}},\ }\href {\doibase 10.1063/1.3518900}
  {\bibfield  {journal} {\bibinfo  {journal} {AIP Conference Proceedings}\
  }\textbf {\bibinfo {volume} {1297}},\ \bibinfo {pages} {135} (\bibinfo {year}
  {2010})}\BibitemShut {NoStop}%
\bibitem [{\citenamefont {Bauer}\ \emph {et~al.}(2011)\citenamefont {Bauer}
  \emph {et~al.}}]{alps}%
  \BibitemOpen
  \bibfield  {author} {\bibinfo {author} {\bibfnamefont {B.}~\bibnamefont
  {Bauer}} \emph {et~al.},\ }\bibfield  {title} {\bibinfo {title} {\emph {The
  ALPS project release 2.0: open source software for strongly correlated
  systems}},\ }\href {\doibase 10.1088/1742-5468/2011/05/P05001} {\bibfield
  {journal} {\bibinfo  {journal} {J. Stat. Mech. Theor. Exp.}\ }\textbf
  {\bibinfo {volume} {2011}},\ \bibinfo {pages} {P05001} (\bibinfo {year}
  {2011})}\BibitemShut {NoStop}%
\bibitem [{\citenamefont {Houcke}\ \emph {et~al.}(2010)\citenamefont {Houcke},
  \citenamefont {Kozik}, \citenamefont {Prokof’ev},\ and\ \citenamefont
  {Svistunov}}]{VanHoucke2010}%
  \BibitemOpen
  \bibfield  {author} {\bibinfo {author} {\bibfnamefont {K.~V.}\ \bibnamefont
  {Houcke}}, \bibinfo {author} {\bibfnamefont {E.}~\bibnamefont {Kozik}},
  \bibinfo {author} {\bibfnamefont {N.}~\bibnamefont {Prokof’ev}}, \ and\
  \bibinfo {author} {\bibfnamefont {B.}~\bibnamefont {Svistunov}},\ }\bibfield
  {title} {\bibinfo {title} {\emph {Diagrammatic Monte Carlo}},\ }\href
  {\doibase 10.1016/j.phpro.2010.09.034} {\bibfield  {journal} {\bibinfo
  {journal} {Physics Procedia}\ }\textbf {\bibinfo {volume} {6}},\ \bibinfo
  {pages} {95 } (\bibinfo {year} {2010})}\BibitemShut {NoStop}%
\bibitem [{\citenamefont {Schollw{\"o}ck}(2005)}]{schollwock1}%
  \BibitemOpen
  \bibfield  {author} {\bibinfo {author} {\bibfnamefont {U.}~\bibnamefont
  {Schollw{\"o}ck}},\ }\bibfield  {title} {\bibinfo {title} {\emph {The
  density-matrix renormalization group}},\ }\href {\doibase
  10.1103/RevModPhys.77.259} {\bibfield  {journal} {\bibinfo  {journal} {Rev.
  Mod. Phys.}\ }\textbf {\bibinfo {volume} {77}},\ \bibinfo {pages} {259}
  (\bibinfo {year} {2005})}\BibitemShut {NoStop}%
\bibitem [{\citenamefont {Schollw{\"o}ck}(2011)}]{schollwock2}%
  \BibitemOpen
  \bibfield  {author} {\bibinfo {author} {\bibfnamefont {U.}~\bibnamefont
  {Schollw{\"o}ck}},\ }\bibfield  {title} {\bibinfo {title} {\emph {The
  density-matrix renormalization group in the age of matrix product states}},\
  }\href {\doibase 10.1016/j.aop.2010.09.012} {\bibfield  {journal} {\bibinfo
  {journal} {Ann. Phys.}\ }\textbf {\bibinfo {volume} {326}},\ \bibinfo {pages}
  {96 } (\bibinfo {year} {2011})}\BibitemShut {NoStop}%
\bibitem [{\citenamefont {Coester}(1958)}]{Coester1958}%
  \BibitemOpen
  \bibfield  {author} {\bibinfo {author} {\bibfnamefont {F.}~\bibnamefont
  {Coester}},\ }\bibfield  {title} {\bibinfo {title} {\emph {{Bound states of a
  many-particle system}}},\ }\href {\doibase 10.1016/0029-5582(58)90280-3}
  {\bibfield  {journal} {\bibinfo  {journal} {Nucl. Phys.}\ }\textbf {\bibinfo
  {volume} {7}},\ \bibinfo {pages} {421} (\bibinfo {year} {1958})}\BibitemShut
  {NoStop}%
\bibitem [{\citenamefont {{\v{C}}{\'\i}{\v{z}}ek}(1966)}]{Cizek1966}%
  \BibitemOpen
  \bibfield  {author} {\bibinfo {author} {\bibfnamefont {J.}~\bibnamefont
  {{\v{C}}{\'\i}{\v{z}}ek}},\ }\bibfield  {title} {\bibinfo {title} {\emph {On
  the correlation problem in atomic and molecular systems. Calculation of
  wavefunction components in Ursell-type expansion using quantum-field
  theoretical methods}},\ }\href {\doibase 10.1063/1.1727484} {\bibfield
  {journal} {\bibinfo  {journal} {J. Chem. Phys.}\ }\textbf {\bibinfo {volume}
  {45}},\ \bibinfo {pages} {4256} (\bibinfo {year} {1966})}\BibitemShut
  {NoStop}%
\bibitem [{\citenamefont {\v{C}\'{\i}\v{z}ek}(1969)}]{Cizek1969}%
  \BibitemOpen
  \bibfield  {author} {\bibinfo {author} {\bibfnamefont {J.}~\bibnamefont
  {\v{C}\'{\i}\v{z}ek}},\ }\bibfield  {title} {\bibinfo {title} {\emph {{On the
  Use of the Cluster Expansion and the Technique of Diagrams in Calculations of
  Correlation Effects in Atoms and Molecules}}},\ }\href {\doibase
  10.1002/9780470143599.ch2} {\bibfield  {journal} {\bibinfo  {journal} {Adv.
  Chem. Phys.}\ }\textbf {\bibinfo {volume} {14}},\ \bibinfo {pages} {35}
  (\bibinfo {year} {1969})}\BibitemShut {NoStop}%
\bibitem [{\citenamefont {{\v{C}}{\'\i}{\v{z}}ek}\ and\ \citenamefont
  {Paldus}(1971)}]{Cizek1971}%
  \BibitemOpen
  \bibfield  {author} {\bibinfo {author} {\bibfnamefont {J.}~\bibnamefont
  {{\v{C}}{\'\i}{\v{z}}ek}}\ and\ \bibinfo {author} {\bibfnamefont
  {J.}~\bibnamefont {Paldus}},\ }\bibfield  {title} {\bibinfo {title} {\emph
  {Correlation problems in atomic and molecular systems III. Rederivation of
  the coupled-pair many-electron theory using the traditional quantum chemical
  methodst}},\ }\href {\doibase 10.1002/qua.560050402} {\bibfield  {journal}
  {\bibinfo  {journal} {Int. J. Quant. Chem.}\ }\textbf {\bibinfo {volume}
  {5}},\ \bibinfo {pages} {359} (\bibinfo {year} {1971})}\BibitemShut {NoStop}%
\bibitem [{\citenamefont {Bartlett}(1981)}]{Bartlett1981}%
  \BibitemOpen
  \bibfield  {author} {\bibinfo {author} {\bibfnamefont {R.~J.}\ \bibnamefont
  {Bartlett}},\ }\bibfield  {title} {\bibinfo {title} {\emph {{Many-Body
  Perturbation Theory and Coupled Cluster Theory for Electron Correlation in
  Molecules}}},\ }\href {\doibase 10.1146/annurev.pc.32.100181.002043}
  {\bibfield  {journal} {\bibinfo  {journal} {Annu. Rev. Phys. Chem.}\ }\textbf
  {\bibinfo {volume} {32}},\ \bibinfo {pages} {359} (\bibinfo {year}
  {1981})}\BibitemShut {NoStop}%
\bibitem [{\citenamefont {Bartlett}(1989)}]{Bartlett1989}%
  \BibitemOpen
  \bibfield  {author} {\bibinfo {author} {\bibfnamefont {R.~J.}\ \bibnamefont
  {Bartlett}},\ }\bibfield  {title} {\bibinfo {title} {\emph {Coupled-cluster
  approach to molecular structure and spectra: a step toward predictive quantum
  chemistry}},\ }\href {\doibase 10.1021/j100342a008} {\bibfield  {journal}
  {\bibinfo  {journal} {J. Phys. Chem.}\ }\textbf {\bibinfo {volume} {93}},\
  \bibinfo {pages} {1697} (\bibinfo {year} {1989})}\BibitemShut {NoStop}%
\bibitem [{\citenamefont {Bishop}(1991)}]{Bishop1991}%
  \BibitemOpen
  \bibfield  {author} {\bibinfo {author} {\bibfnamefont {R.}~\bibnamefont
  {Bishop}},\ }\bibfield  {title} {\bibinfo {title} {\emph {An overview of
  coupled cluster theory and its applications in physics}},\ }\href {\doibase
  10.1007/BF01119617} {\bibfield  {journal} {\bibinfo  {journal} {Theor. Chem.
  Acc.}\ }\textbf {\bibinfo {volume} {80}},\ \bibinfo {pages} {95} (\bibinfo
  {year} {1991})}\BibitemShut {NoStop}%
\bibitem [{\citenamefont {Paldus}\ and\ \citenamefont {Li}(1999)}]{Paldus1999}%
  \BibitemOpen
  \bibfield  {author} {\bibinfo {author} {\bibfnamefont {J.}~\bibnamefont
  {Paldus}}\ and\ \bibinfo {author} {\bibfnamefont {X.}~\bibnamefont {Li}},\
  }\bibfield  {title} {\bibinfo {title} {\emph {A Critical Assessment of
  Coupled Cluster Method in Quantum Chemistry}},\ }\href {\doibase
  10.1002/9780470141694.ch1} {\bibfield  {journal} {\bibinfo  {journal} {Adv.
  Chem. Phys.}\ }\textbf {\bibinfo {volume} {110}},\ \bibinfo {pages} {1}
  (\bibinfo {year} {1999})}\BibitemShut {NoStop}%
\bibitem [{\citenamefont {Bartlett}\ and\ \citenamefont
  {Musia{\l}}(2007)}]{Musial2007}%
  \BibitemOpen
  \bibfield  {author} {\bibinfo {author} {\bibfnamefont {R.~J.}\ \bibnamefont
  {Bartlett}}\ and\ \bibinfo {author} {\bibfnamefont {M.}~\bibnamefont
  {Musia{\l}}},\ }\bibfield  {title} {\bibinfo {title} {\emph {Coupled-cluster
  theory in quantum chemistry}},\ }\href {\doibase 10.1103/RevModPhys.79.291}
  {\bibfield  {journal} {\bibinfo  {journal} {Rev. Mod. Phys.}\ }\textbf
  {\bibinfo {volume} {79}},\ \bibinfo {pages} {291} (\bibinfo {year}
  {2007})}\BibitemShut {NoStop}%
\bibitem [{\citenamefont {Lyakh}\ \emph {et~al.}(2012)\citenamefont {Lyakh},
  \citenamefont {Musia{\l}}, \citenamefont {Lotrich},\ and\ \citenamefont
  {Bartlett}}]{Lyakh2012}%
  \BibitemOpen
  \bibfield  {author} {\bibinfo {author} {\bibfnamefont {D.~I.}\ \bibnamefont
  {Lyakh}}, \bibinfo {author} {\bibfnamefont {M.}~\bibnamefont {Musia{\l}}},
  \bibinfo {author} {\bibfnamefont {V.~F.}\ \bibnamefont {Lotrich}}, \ and\
  \bibinfo {author} {\bibfnamefont {R.~J.}\ \bibnamefont {Bartlett}},\
  }\bibfield  {title} {\bibinfo {title} {\emph {{Multireference nature of
  chemistry: the coupled-cluster view.}}},\ }\href {\doibase 10.1021/cr2001417}
  {\bibfield  {journal} {\bibinfo  {journal} {Chem. Rev.}\ }\textbf {\bibinfo
  {volume} {112}},\ \bibinfo {pages} {182} (\bibinfo {year}
  {2012})}\BibitemShut {NoStop}%
\bibitem [{\citenamefont {Bishop}\ \emph {et~al.}(1994)\citenamefont {Bishop},
  \citenamefont {Hale},\ and\ \citenamefont {Xian}}]{Bishop1994}%
  \BibitemOpen
  \bibfield  {author} {\bibinfo {author} {\bibfnamefont {R.~F.}\ \bibnamefont
  {Bishop}}, \bibinfo {author} {\bibfnamefont {R.~G.}\ \bibnamefont {Hale}}, \
  and\ \bibinfo {author} {\bibfnamefont {Y.}~\bibnamefont {Xian}},\ }\bibfield
  {title} {\bibinfo {title} {\emph {Systematic Inclusion of High-Order
  Multispin Correlations for the Spin-1/2 $\mathrm{XXZ}$ Models}},\ }\href
  {\doibase 10.1103/PhysRevLett.73.3157} {\bibfield  {journal} {\bibinfo
  {journal} {Phys. Rev. Lett.}\ }\textbf {\bibinfo {volume} {73}},\ \bibinfo
  {pages} {3157} (\bibinfo {year} {1994})}\BibitemShut {NoStop}%
\bibitem [{\citenamefont {Bishop}\ and\ \citenamefont {Li}(2011)}]{Bishop2011}%
  \BibitemOpen
  \bibfield  {author} {\bibinfo {author} {\bibfnamefont {R.~F.}\ \bibnamefont
  {Bishop}}\ and\ \bibinfo {author} {\bibfnamefont {P.~H.~Y.}\ \bibnamefont
  {Li}},\ }\bibfield  {title} {\bibinfo {title} {\emph {Coupled-cluster method:
  A lattice-path-based subsystem approximation scheme for quantum lattice
  models}},\ }\href {\doibase 10.1103/PhysRevA.83.042111} {\bibfield  {journal}
  {\bibinfo  {journal} {Phys. Rev. A}\ }\textbf {\bibinfo {volume} {83}},\
  \bibinfo {pages} {042111} (\bibinfo {year} {2011})}\BibitemShut {NoStop}%
\bibitem [{\citenamefont {Cederbaum}\ \emph {et~al.}(2006)\citenamefont
  {Cederbaum}, \citenamefont {Alon},\ and\ \citenamefont
  {Streltsov}}]{Cederbaum2006}%
  \BibitemOpen
  \bibfield  {author} {\bibinfo {author} {\bibfnamefont {L.~S.}\ \bibnamefont
  {Cederbaum}}, \bibinfo {author} {\bibfnamefont {O.~E.}\ \bibnamefont {Alon}},
  \ and\ \bibinfo {author} {\bibfnamefont {A.~I.}\ \bibnamefont {Streltsov}},\
  }\bibfield  {title} {\bibinfo {title} {\emph {Coupled-cluster theory for
  systems of bosons in external traps}},\ }\href {\doibase
  10.1103/PhysRevA.73.043609} {\bibfield  {journal} {\bibinfo  {journal} {Phys.
  Rev. A}\ }\textbf {\bibinfo {volume} {73}},\ \bibinfo {pages} {043609}
  (\bibinfo {year} {2006})}\BibitemShut {NoStop}%
\bibitem [{\citenamefont {Alon}\ \emph {et~al.}(2006)\citenamefont {Alon},
  \citenamefont {Streltsov},\ and\ \citenamefont {Cederbaum}}]{Alon2006}%
  \BibitemOpen
  \bibfield  {author} {\bibinfo {author} {\bibfnamefont {O.~E.}\ \bibnamefont
  {Alon}}, \bibinfo {author} {\bibfnamefont {A.~I.}\ \bibnamefont {Streltsov}},
  \ and\ \bibinfo {author} {\bibfnamefont {L.~S.}\ \bibnamefont {Cederbaum}},\
  }\bibfield  {title} {\bibinfo {title} {\emph {Coupled-cluster theory for
  bosons in rings and optical lattices}},\ }\href {\doibase
  10.1016/j.theochem.2006.05.026} {\bibfield  {journal} {\bibinfo  {journal}
  {J. Mol. Struct. THEOCHEM}\ }\textbf {\bibinfo {volume} {768}},\ \bibinfo
  {pages} {151} (\bibinfo {year} {2006})}\BibitemShut {NoStop}%
\bibitem [{\citenamefont {Hofmann}\ \emph {et~al.}(2015)\citenamefont
  {Hofmann}, \citenamefont {Lobos},\ and\ \citenamefont
  {Galitski}}]{Hofmann2015}%
  \BibitemOpen
  \bibfield  {author} {\bibinfo {author} {\bibfnamefont {J.}~\bibnamefont
  {Hofmann}}, \bibinfo {author} {\bibfnamefont {A.~M.}\ \bibnamefont {Lobos}},
  \ and\ \bibinfo {author} {\bibfnamefont {V.}~\bibnamefont {Galitski}},\
  }\bibfield  {title} {\bibinfo {title} {\emph {Parity effect and few-to-many
  particle crossover in a mesoscopic Fermi gas}},\ }\href
  {http://arxiv.org/abs/1508.05947} {\bibfield  {journal} {\bibinfo  {journal}
  {arXiv:1508.05947}\ } (\bibinfo {year} {2015})}\BibitemShut {NoStop}%
\bibitem [{\citenamefont {Busch}\ \emph {et~al.}(1998)\citenamefont {Busch},
  \citenamefont {Englert}, \citenamefont {Rza{\.z}ewski},\ and\ \citenamefont
  {Wilkens}}]{Busch1998}%
  \BibitemOpen
  \bibfield  {author} {\bibinfo {author} {\bibfnamefont {T.}~\bibnamefont
  {Busch}}, \bibinfo {author} {\bibfnamefont {B.-G.}\ \bibnamefont {Englert}},
  \bibinfo {author} {\bibfnamefont {K.}~\bibnamefont {Rza{\.z}ewski}}, \ and\
  \bibinfo {author} {\bibfnamefont {M.}~\bibnamefont {Wilkens}},\ }\bibfield
  {title} {\bibinfo {title} {\emph {Two Cold Atoms in a Harmonic Trap}},\
  }\href {\doibase 10.1023/A:1018705520999} {\bibfield  {journal} {\bibinfo
  {journal} {Found. Phys.}\ }\textbf {\bibinfo {volume} {28}},\ \bibinfo
  {pages} {549} (\bibinfo {year} {1998})}\BibitemShut {NoStop}%
\bibitem [{\citenamefont {Sherrill}\ and\ \citenamefont
  {Schaefer}(1999)}]{Sherrill1999}%
  \BibitemOpen
  \bibfield  {author} {\bibinfo {author} {\bibfnamefont {C.~D.}\ \bibnamefont
  {Sherrill}}\ and\ \bibinfo {author} {\bibfnamefont {H.~F.}\ \bibnamefont
  {Schaefer}},\ }\bibfield  {title} {\bibinfo {title} {\emph {{The
  Configuration Interaction Method: Advances in Highly Correlated
  Approaches}}},\ }\href {\doibase 10.1016/S0065-3276(08)60532-8} {\bibfield
  {journal} {\bibinfo  {journal} {Adv. Quantum Chem.}\ }\textbf {\bibinfo
  {volume} {34}},\ \bibinfo {pages} {143} (\bibinfo {year} {1999})}\BibitemShut
  {NoStop}%
\bibitem [{\citenamefont {Olsen}\ \emph {et~al.}(1988)\citenamefont {Olsen},
  \citenamefont {Roos}, \citenamefont {Jo̸rgensen},\ and\ \citenamefont
  {Jensen}}]{Olsen1988}%
  \BibitemOpen
  \bibfield  {author} {\bibinfo {author} {\bibfnamefont {J.}~\bibnamefont
  {Olsen}}, \bibinfo {author} {\bibfnamefont {B.~O.}\ \bibnamefont {Roos}},
  \bibinfo {author} {\bibfnamefont {P.}~\bibnamefont {Jo̸rgensen}}, \ and\
  \bibinfo {author} {\bibfnamefont {H.~J.~A.}\ \bibnamefont {Jensen}},\
  }\bibfield  {title} {\bibinfo {title} {\emph {{Determinant based
  configuration interaction algorithms for complete and restricted
  configuration interaction spaces}}},\ }\href {\doibase 10.1063/1.455063}
  {\bibfield  {journal} {\bibinfo  {journal} {J. Chem. Phys.}\ }\textbf
  {\bibinfo {volume} {89}},\ \bibinfo {pages} {2185} (\bibinfo {year}
  {1988})}\BibitemShut {NoStop}%
\bibitem [{\citenamefont {Coester}\ and\ \citenamefont
  {K{\"u}mmel}(1960)}]{Coester1960}%
  \BibitemOpen
  \bibfield  {author} {\bibinfo {author} {\bibfnamefont {F.}~\bibnamefont
  {Coester}}\ and\ \bibinfo {author} {\bibfnamefont {H.}~\bibnamefont
  {K{\"u}mmel}},\ }\bibfield  {title} {\bibinfo {title} {\emph {Short-range
  correlations in nuclear wave functions}},\ }\href {\doibase
  10.1016/0029-5582(60)90140-1} {\bibfield  {journal} {\bibinfo  {journal}
  {Nucl. Phys.}\ }\textbf {\bibinfo {volume} {17}},\ \bibinfo {pages} {477}
  (\bibinfo {year} {1960})}\BibitemShut {NoStop}%
\bibitem [{\citenamefont {Raghavachari}\ \emph {et~al.}(1989)\citenamefont
  {Raghavachari}, \citenamefont {Trucks}, \citenamefont {Pople},\ and\
  \citenamefont {Head-Gordon}}]{Raghavachari1989}%
  \BibitemOpen
  \bibfield  {author} {\bibinfo {author} {\bibfnamefont {K.}~\bibnamefont
  {Raghavachari}}, \bibinfo {author} {\bibfnamefont {G.~W.}\ \bibnamefont
  {Trucks}}, \bibinfo {author} {\bibfnamefont {J.~A.}\ \bibnamefont {Pople}}, \
  and\ \bibinfo {author} {\bibfnamefont {M.}~\bibnamefont {Head-Gordon}},\
  }\bibfield  {title} {\bibinfo {title} {\emph {{Reprint of: A fifth-order
  perturbation comparison of electron correlation theories}}},\ }\href
  {\doibase 10.1016/j.cplett.2013.08.064} {\bibfield  {journal} {\bibinfo
  {journal} {Chem. Phys. Lett.}\ }\textbf {\bibinfo {volume} {589}},\ \bibinfo
  {pages} {37} (\bibinfo {year} {1989})}\BibitemShut {NoStop}%
\bibitem [{\citenamefont {{Grining}}\ \emph {et~al.}(2015)\citenamefont
  {{Grining}}, \citenamefont {{Tomza}}, \citenamefont {{Lesiuk}}, \citenamefont
  {{Przybytek}}, \citenamefont {{Musia{\l}}}, \citenamefont {{Massignan}},
  \citenamefont {{Lewenstein}},\ and\ \citenamefont
  {{Moszynski}}}]{Grining2015}%
  \BibitemOpen
  \bibfield  {author} {\bibinfo {author} {\bibfnamefont {T.}~\bibnamefont
  {{Grining}}}, \bibinfo {author} {\bibfnamefont {M.}~\bibnamefont {{Tomza}}},
  \bibinfo {author} {\bibfnamefont {M.}~\bibnamefont {{Lesiuk}}}, \bibinfo
  {author} {\bibfnamefont {M.}~\bibnamefont {{Przybytek}}}, \bibinfo {author}
  {\bibfnamefont {M.}~\bibnamefont {{Musia{\l}}}}, \bibinfo {author}
  {\bibfnamefont {P.}~\bibnamefont {{Massignan}}}, \bibinfo {author}
  {\bibfnamefont {M.}~\bibnamefont {{Lewenstein}}}, \ and\ \bibinfo {author}
  {\bibfnamefont {R.}~\bibnamefont {{Moszynski}}},\ }\bibfield  {title}
  {\bibinfo {title} {\emph {Many interacting fermions in a one-dimensional
  harmonic trap: a quantum-chemical treatment}},\ }\href
  {http://stacks.iop.org/1367-2630/17/i=11/a=115001} {\bibfield  {journal}
  {\bibinfo  {journal} {New J. Phys.}\ }\textbf {\bibinfo {volume} {17}},\
  \bibinfo {pages} {115001} (\bibinfo {year} {2015})}\BibitemShut {NoStop}%
\bibitem [{\citenamefont {Przybytek}()}]{HECTOR}%
  \BibitemOpen
  \bibfield  {author} {\bibinfo {author} {\bibfnamefont {M.}~\bibnamefont
  {Przybytek}},\ }\href@noop {} {\bibinfo {title} {\emph {{FCI program HECTOR;
  University of Warsaw, (unpublished)}}}}\BibitemShut {NoStop}%
\bibitem [{\citenamefont {Stanton}\ \emph {et~al.}()\citenamefont {Stanton},
  \citenamefont {Gauss}, \citenamefont {Watts}, \citenamefont {Lauderdale},\
  and\ \citenamefont {Bartlett}}]{ACESII}%
  \BibitemOpen
  \bibfield  {author} {\bibinfo {author} {\bibfnamefont {J.~F.}\ \bibnamefont
  {Stanton}}, \bibinfo {author} {\bibfnamefont {J.}~\bibnamefont {Gauss}},
  \bibinfo {author} {\bibfnamefont {J.~D.}\ \bibnamefont {Watts}}, \bibinfo
  {author} {\bibfnamefont {W.~J.}\ \bibnamefont {Lauderdale}}, \ and\ \bibinfo
  {author} {\bibfnamefont {R.~J.}\ \bibnamefont {Bartlett}},\ }\href@noop {}
  {\bibinfo {title} {\emph {{ACES II Program System Release 2.0 QTP; University
  of Florida: Gainesville, FL, 1994.}}}}\BibitemShut {Stop}%
\end{thebibliography}%

\end{document}